\documentclass{aa}
\usepackage{epsfig}
\usepackage{lscape}
\usepackage{aalongtable}
\usepackage{natbib}
\usepackage{graphics}
\usepackage{color}
\usepackage{ulem}    
\usepackage{times}
\begin{document}
\title{VLT spectroscopy of low-metallicity emission-line galaxies: 
abundance patterns and abundance 
discrepancies\thanks{Based on 
observations
collected at the European Southern Observatory, Chile. List of programs is
shown in Table~\ref{tab1}.}
\thanks{ Tables~\ref{tab2} - \ref{tab7} and \ref{tab10} and Figures~\ref{fig1} - \ref{fig4} 
are only available in electronic form in the online
edition.}
}
\author{N. G.\ Guseva \inst{1,2}
\and Y. I.\ Izotov \inst{1,2,3}
\and G.\ Stasi\'nska \inst{3}
\and K. J.\ Fricke \inst{1,4}
\and C.\ Henkel \inst{1}
\and P.\ Papaderos \inst{5,6}}
\offprints{N.G. Guseva, guseva@mao.kiev.ua}
\institute{Max-Planck-Institute for Radioastronomy, Auf dem H\"ugel 69,
                     53121 Bonn, Germany
\and 
                     Main Astronomical Observatory,
                     Ukrainian National Academy of Sciences,
                     Zabolotnoho 27, Kyiv 03680,  Ukraine
\and
                     LUTH, Observatoire de Meudon, F-92195 Meudon Cedex, France
\and
                     Institute for Astrophysics, University of 
                     G\"ottingen, Friedrich-Hund-Platz 1, 
                     37077 G\"ottingen, Germany
\and                 
                     Centro de Astrof\'{\i}sica da Universidade do Porto,
                     Rua das Esteral, 4150-762 porto, Portugal 
\and
                     Department of Astronomy, Oskar Klein Centre,
                     Stockholm University, SE-106 91 Stockholm, Sweden
}

\date{Received \hskip 2cm; Accepted}

\abstract
{We present deep spectroscopy of a large sample of low-metallicity
emission-line galaxies.
}
{The main goal of this study is to derive element abundances
in these low-metallicity galaxies.
} 
{We analyze 121 VLT spectra of H {{\sc ii}} regions in 46 low-metallicity 
emission-line galaxies. 83 of these spectra are archival VLT/FORS1+UVES spectra
of H {{\sc ii}} regions in 31 
low-metallicity emission-line galaxies that are 
studied for the first time with standard direct methods
to determine the electron temperatures, the electron number densities, 
and the chemical abundances. 
}
{The oxygen abundance of the sample lies in the range 
12 + log O/H = 7.2 -- 8.4. 
We confirm previous findings that
Ne/O increases with increasing oxygen abundance, likely because of a higher
depletion of oxygen in higher-metallicity galaxies. 
The Fe/O ratio decreases from roughly solar at the lowest 
metallicities to about 
one tenth of solar, indicating that the degree of depletion of iron into dust 
grains depends on metallicity. The N/O ratio in extremely 
low-metallicity galaxies 
with 12 + log O/H $<$ 7.5 shows a slight increase  with decreasing 
oxygen abundance, 
which could be the signature of enhanced production of primary nitrogen by 
rapidly rotating stars at low metallicity.
We present the first empirical relation between the electron temperature 
derived from 
[S {{\sc iii}}]$\lambda$6312/$\lambda$9069 or 
[N {{\sc ii}}]$\lambda$5755/$\lambda$6583  and  the one derived from 
[O {{\sc iii}}]$\lambda$4363/$\lambda$(4959+5007) in low-metallicity galaxies.
We also present an empirical relation between $t_e$ derived from 
[O {{\sc ii}}]$\lambda$3727/($\lambda$7320 $+$ $\lambda$7330) or 
[S {{\sc ii}}]$\lambda$4068/($\lambda$6717 $+$ $\lambda$6730) and 
[O {{\sc iii}}]$\lambda$4363/$\lambda$(4959+5007).
The electron number densities  $N_e$(Cl {{\sc iii}}) and 
$N_e$(Ar {{\sc iv}}) were derived in a number of objects and are found to be 
higher than $N_e$(O {{\sc ii}}) and $N_e$(S {{\sc ii}}). This has potential 
implications for the derivation of the pregalactic helium abundance. In a 
number of objects, the abundances of C$^{++}$ and O$^{++}$ could be derived 
from recombination lines. Our study confirms the discrepancy between abundances
found from recombination lines (RLs) 
and collisionally excited lines (CELs) and that C/O increases with O/H.
}
{}

\keywords{galaxies: fundamental parameters --
galaxies: starburst -- galaxies: abundances -- galaxies: ISM}
\titlerunning{Low-metallicity ELGs: deep VLT/FORS+UVES spectroscopy}
\authorrunning{N.G.Guseva et al.}
\maketitle

\section{Introduction \label{intro}}

\begin{table*}
  \caption{H {\sc ii} regions and galaxies observed with the VLT \label{tab1}}
\begin{tabular}{lrrll} \hline
 Name        & R.A. (J2000.0) & Dec. (J2000.0)  & Instrument &ESO Program \\ 
\hline \hline
UM 254           &00:31:34.2      &$-$02:09:18.0    & UVES               &71.B-0055(A) \\
UM 283           &00:51:49.6      &$+$00:33:55.4    & UVES               &71.B-0055(A),70.B-0717(A) \\
NGC 346          &00:59:06.0      &$-$72:10:14.5    & FORS low+medium resolution  &69.C-0203(A) \\
NGC 456          &01:13:53.3      &$-$73:17:49.8    & FORS low+medium resolution  &69.C-0203(A) \\
UM 133           &01:44:41.3      &$+$04:53:24.3    & UVES               &68.B-0310(A),70.B-0717(A) \\
UM 382           &01:58:09.5      &$-$00:06:37.1    & UVES               &70.B-0717(A) \\
UM 408           &02:11:23.5      &$+$02:20:32.0    & UVES               &70.B-0717(A) \\
UM 417           &02:19:30.2      &$-$00:59:16.2    & UVES               &70.B-0717(A) \\ 
UM 420           &02:20:54.5      &$+$00:33:24.0    & FORS low+medium resolution  &69.C-0203(A) \\
MRK 600          &02:51:04.5      &$+$04:27:15.0    & UVES               &70.B-0717(A) \\
CAM 0357$-$3915  &03:59:08.9      &$-$39:06:25.0    & FORS low+medium resolution  &69.C-0203(A) \\
TOL 0513$-$393   &05:15:19.8      &$-$39:17:41.0    & FORS low+medium resolution  &69.C-0203(A) \\
TOL 0618$-$402   &06:20:02.5      &$-$40:18:09.0    & UVES               &70.B-0717(A) \\
HE 2-10          &08:36:15.2      &$-$26:24:34.3    & UVES               &073.B-0283(A), 081.C-0113(A) \\
NGC 3125         &10:06:33.3      &$-$29:56:07.5    & UVES               &081.C-0113(A) \\  
MRK 1259         &10:38:33.6      &$-$07:10:14.3    & UVES               &073.B-0283(A) \\ 
MRK 1271         &10:56:09.1      &$+$06:10:22.0    & UVES               &081.C-0113(A) \\
POX 4            &11:51:11.5      &$-$20:35:58.8    & UVES               &081.C-0113(A) \\
TOL 1214$-$277   &12:17:18.7      &$-$28:02:21.3    & UVES, FORS medium resolution &69.D-0174(A), 65.N-0642(A) \\
TOL 65           &12:25:47.9      &$-$36:13:45.9    & FORS medium resolution      &65.N-0642(A) \\
J1253$-$0312     &12:53:05.9      &$-$03:12:58.5    & UVES               &081.C-0113(A) \\ 
NGC 5253         &13:39:55.9      &$-$31:38:24.7    & UVES               &70.C-0008(A), 073.B-0283(A) \\
TOL 89           &14:01:20.0      &$-$33:04:11.2    & UVES               &073.B-0283(A) \\
NGC 5408         &14:03:18.4      &$-$41:22:50.1    & UVES               &081.C-0113(A) \\ 
TOL 1457$-$262   &15:00:28.9      &$-$26:26:50.0    & UVES               &081.C-0113(A) \\
TOL 1924$-$416   &19:27:58.1      &$-$41:34:31.2    & UVES               &081.C-0113(A) \\
NGC 6822 V       &19:44:52.7      &$-$14:43:09.8    & UVES               &081.C-0113(A) \\
NGC 6822 V+X     &19:44:54.5      &$-$14:43:02.9    & FORS low+medium resolution  &69.C-0203(A) \\
TOL 2138$-$405   &21:41:21.8      &$-$40:19:06.0    & FORS low+medium resolution  &69.C-0203(A) \\
TOL 2146$-$391   &21:49:48.2      &$-$38:54:09.0    & FORS low+medium resolution  &69.C-0203(A) \\
PHL 293B         &22:30:36.6      &$-$00:06:37.3    & UVES               &70.B-0717(A) \\
TOL 2240$-$384   &22:43:33.0      &$-$38:10:55.2    & FORS low+medium resolution  &69.C-0203(A) \\
UM 160           &23:24:22.0      &$-$00:06:21.2    & FORS low+medium resolution  &69.C-0203(A) \\
\hline \hline
\end{tabular}
\end{table*}

 Comprehensive studies of low-metallicity 
emission-line galaxies in 
the nearby universe are essential to investigate star
formation and galaxy evolution with chemical conditions 
close to those in the early universe.
 In particular, studies of abundance ratios and their dependencies on 
metallicity in low-metallicity galaxies 
are important for studying the early chemical evolution 
of galaxies and the nucleosynthesis of 
massive stars in an environment characterized by a nearly 
pristine chemical composition. 
Improved statistics of low-metallicity galaxies, observed with high accuracy,
are also important to put observational
constraints on the primordial helium abundance, 
which is a key parameter for testing cosmological models.

We here continue our study of 
nearby (redshift $z$ $\la$ 0.1)  
star-forming galaxies 
with measured intensities of the [O {\sc iii}] $\lambda$4363 emission line 
\citep{G2006, Iz06, Pap2006a, G2007, Pap2008}.
This line is important to \textit{directly}
determine the physical conditions and chemical abundances of 
the ionized interstellar medium.

 There were many spectroscopic surveys in the past
aiming to study element abundances of nearby low-metallicity emission-line 
galaxies \citep[e.g. ][]{KS83,C86,T91,M94,MS02}. 
However, most of these spectroscopic data are not suitable for our goals
for various reasons, such as an insufficient quality of the data, 
the nonlinearity of the detectors in observations before the 1990ies, 
and the limited wavelength coverage.

In this paper, we consider archival data from the  8.2\,m 
Very Large Telescope (VLT) obtained in the period 2000 - 2008 
with the spectrographs FORS1 
and UVES, which allow us to compile a sample of 83 high-quality spectra 
of metal-poor  H {\sc ii} regions. 
 We merge this sample with 15 archival VLT/FORS1+UVES
spectra of two BCD galaxies, SBS 0335--052E and SBS 0335--052W
\citep{IGFP2009}, and VLT/FORS2 spectra of 
23 H {\sc ii} regions in 12 low-metallicity emission-line galaxies  
from \citet{G2009}.
 We thus obtain a
large VLT sample of 121 H {\sc ii} regions, 
which includes to our knowledge all high-quality spectra 
from low-metallicity objects observed with the VLT/UVES+FORS instruments.
They are used to study the
abundance patterns of low-metallicity emission-line galaxies. 

 For comparison reasons we supplemented this sample with 109 low-metallicity
emission-line galaxies from
the HeBCD sample that were observed with different telescopes and collected 
by \citet{IT04a} and \citet{ISGT2004} for the study of the primordial He 
abundance. 
  The total sample consists of objects with low extinction 
and in general with a high equivalent width EW(H$\beta$) of the H$\beta$
emission line, which suggests a young age of the respective starburst. 
It also includes
almost all most metal-deficient galaxies known, which allows 
 for the deepest insights into
the abundance pattern  of the lowest-metallicity galaxies
available in the local Universe. Additionally, we compare abundance patterns
of this sample with those of the galaxies from the Data Release 3 (DR3) of 
the Sloan Digital Sky survey (SDSS) by \citet{Iz06} with lower quality spectra 
but higher metallicity, thus extending the range of oxygen abundances to
higher values. H {\sc ii} regions in the above samples are selected mainly
because of their high apparent brightness to obtain high 
signal-to-noise ratio spectra, and not considering their physical properties,
such as the H$\beta$ luminosity. However, these differing selection criteria 
do not introduce biases. In particular, \citet{I11} have shown that 
the abundances and abundance ratios of luminous compact galaxies selected
from the SDSS by their high H$\beta$ luminosity do not differ from the
samples selected by their apparent brightness. 

 This paper is organized as follows:  in Section 2 we describe observations and data 
reduction. In Section 3 we comment on the physical parameters of the  H {\sc ii} regions. 
In Section 4 we discuss the abundance patterns derived from the analysis of collisionally 
excited lines and present carbon and oxygen abundances obtained from the analysis of recombination lines. 
 We summarize our conclusions in Section 5.

\section{Observations and data reduction}

From the ESO archive we selected VLT spectra of H {\sc ii} regions that 
were obtained with the spectrographs UVES (high resolution)
and FORS1 (low and medium resolution).
The list of the 30 emission-line galaxies to which these  H {\sc ii} 
regions pertain as well as the two H {\sc ii} regions 
NGC 346 and NGC 456 in the Small Magellanic Cloud
is given in Table~\ref{tab1} together with the equatorial coordinates,
the spectrographs used and the identification number of the ESO program. 
  The total number of spectra from these galaxies
available for abundance determinations (and 
listed in Table~\ref{tab8}) is 83 -- larger than the list in 
Table~\ref{tab1}. 
This is because in some galaxies several one-dimensional spectra of
H {\sc ii} regions were extracted from two-dimensional frames. 
One galaxy, Tol1214--277, was observed with
two different spectrographs. Furthermore, the entire objects from the 
ESO program
69.C-0203(A) were observed with low and medium resolution. 
 Four galaxies were observed twice with VLT/UVES within two different ESO 
programmes.  The images of the observed galaxies overlayed by the slit 
positions are shown in Fig. \ref{fig1}.

The raw spectra of the objects from the ESO archive as well as the 
calibration spectra 
 were reduced using IRAF\footnote{IRAF is distributed by the National 
Optical Astronomy 
Observatory, which is operated by the Association of Universities for 
Research in Astronomy, Inc., under cooperative agreement with the National 
Science Foundation.}.
The details of data reduction are described in \citet{TI2005}. 
The two-dimensional spectra were bias subtracted and flat-field corrected. 
We then used the IRAF software routines IDENTIFY, REIDENTIFY, FITCOORD, and
TRANSFORM to perform wavelength
calibration and correction for distortion and tilt for each frame. 
Night sky subtraction was performed using the routine BACKGROUND. The level of
night sky emission was determined from the regions closest to the H {\sc ii}
region which are free of stellar and nebular line emission,
as well as of emission from foreground and background sources.
Exceptions are the extended Small Magellanic Cloud (SMC) 
H {\sc ii} regions NGC 346 and NGC 456, for 
which no night sky subtraction was done.
One-dimensional spectra were then extracted from each two-dimensional 
frame using the APALL routine. Before extraction, distinct two-dimensional 
spectra of the same H {\sc ii} region
were carefully aligned using the spatial locations of the brightest part in
each spectrum, so that spectra were extracted at the same positions in all
subexposures. We have summed the individual spectra 
from each subexposure after removal of the cosmic ray hits.

The high-resolution spectra of H {\sc ii} regions observed with the UVES
spectrograph are shown in Fig.~\ref{fig2}, while low-resolution and 
medium-resolution FORS1 spectra are shown in Figs.~\ref{fig3} and ~\ref{fig4},
respectively. All these figures are available in electronic form only.
 Emission-line fluxes were measured using the IRAF SPLOT routine.
 The line flux errors 
include statistical errors derived with SPLOT
from non-flux-calibrated spectra, in addition to errors introduced
by the absolute flux calibration, which we
set to 1\% of the line fluxes, according to the uncertainties of 
absolute fluxes of relatively bright standard stars 
\citep{Oke1990,Colina1994,Bohlin1996,IT04a}. 
These errors will be later propagated into the calculation of 
the electron temperatures, the electron number densities, and ionic and 
the element abundances.
  Given a function $f$($x,y,...,z$), the uncertainty $\sigma$$(f)$
is calculated as

\begin{equation}
\sigma(f)=\sqrt{\left(\frac{df}{dx}\right)^2\sigma(x)^2 + 
\left(\frac{df}{dy}\right)^2\sigma(y)^2 + ...
+ \left(\frac{df}{dz}\right)^2\sigma(z)^2}.
\label{sigma}
\end{equation}

  These errors do not include 
uncertainties introduced by the standard data reduction.
 In particular, one source of uncertainties comes from the division of 
object and standard star frames by the flat field frame. This is because 
the objects (especially those that are extended and have multiple knots) 
and the spectrophotometric
standard star spectra are located in the frame at slightly different positions 
corresponding to slightly different intensities of the flat field. 
Additionally, atmospheric refraction may 
play a role resulting from the varying inclination of 
the spectra in the frame depending on airmass and position angle.
  Fringes in the red part of spectra introduce an additional source of 
uncertainties. 
  The  errors introduced by the standard data reduction and these fringes 
can be estimated by comparing the line intensities in different 
observations of the same  H {\sc ii} region, although differences in slit 
positions and apertures may also contribute to differences
in line intensities.

The corrections for underlying absorption in hydrogen lines and for 
extinction were performed following the procedure desribed in 
\citet{G2009}.
 We show in Tables~\ref{tab2}, \ref{tab3}, and \ref{tab4}
the extinction-corrected 
emission line fluxes  
relative to the H$\beta$ fluxes
along with the extinction coefficients $C$(H$\beta$), the observed fluxes 
$F$(H$\beta$) of the H$\beta$ emission line, the equivalent widths EW(H$\beta$)
of the H$\beta$ emission line, and the equivalent widths EW(abs) of the hydrogen
absorption lines. All these tables are available in electronic form only.  
Table~\ref{tab2} contains the UVES observations, Table~\ref{tab3} 
the low-resolution, and Table \ref{tab4} 
the medium-resolution FORS observations. 

To constitute a VLT sample, that is as comprehensive
as possible, we added to the data described above 
 15 archival VLT/FORS1+UVES
spectra
from the two BCD galaxies SBS 0335--052E and SBS 0335--052W
studied before by \citet{IGFP2009} and 23 VLT/FORS2 spectra from 12 galaxies 
selected mainly from
Data Release 6 (DR6) of the Sloan Digital Sky Survey (SDSS)
and studied by \citet{G2009}. Our resulting VLT sample thus  contains 
121 spectra 
of H {{\sc ii}} regions from 46 low-metallicity emission-line galaxies.
 Eighty-three of these are archival VLT/FORS1+UVES spectra 
that are analyzed for the first time.
 All these spectra were observed with the same telescope and reduced 
in the same way.

\setcounter{figure}{4}

\begin{figure}[t]
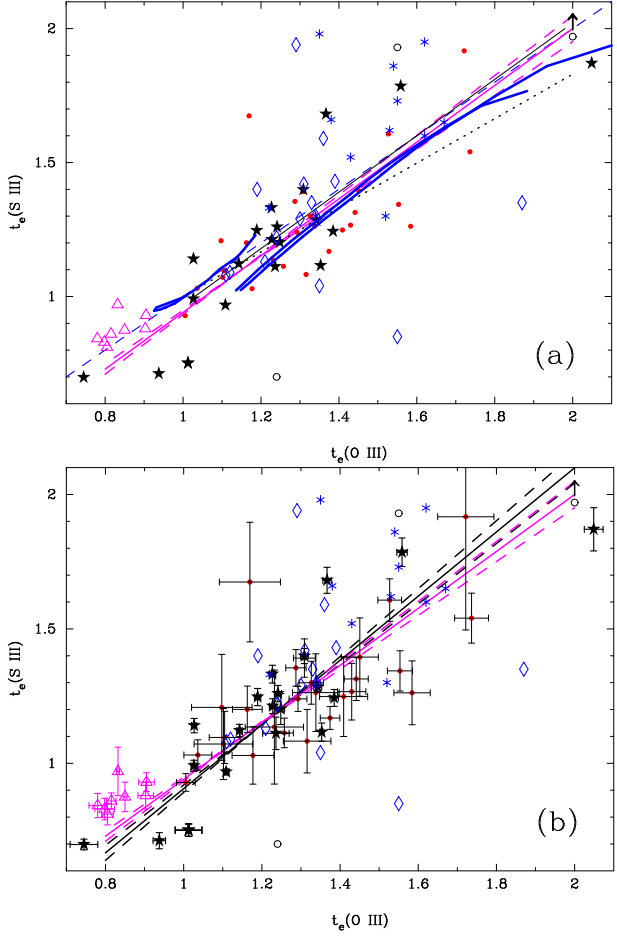

\hspace*{0.0cm}\psfig{figure=16291f5_1.ps,angle=-90,width=8.cm,clip=}
\hspace*{0.0cm}\psfig{figure=16291f5_2.ps,angle=-90,width=8.cm,clip=}
\caption{Comparison of electron temperatures $t_e$(O {{\sc iii}}) 
($t_e$ = 10$^{-4}$ $T_e$)
obtained from [O {{\sc iii}}]$\lambda$4363/($\lambda$4959 $+$ $\lambda$5007)
and $t_e$(S {{\sc iii}}) obtained from 
[S {{\sc iii}}]$\lambda$6312/$\lambda$9069 emission-line ratios.
VLT/UVES data are shown by stars, except for three objects 
with uncertain values, which are shown as open circles 
(see Sect.~\ref{te}).  
   Twenty-seven data points from SDSS DR3 (see text) are shown by red dots. 
 Open blue diamonds, blue asterisks, and purple triangles
correspond to data from  \citet{Kehrig06}, \citet{PerezDiaz_03} and
\citet{Garcia_Esteb2007}, respectively. 
  The dashed blue line in the upper panel connects locations 
of equal temperature. 
 The thick blue lines (also in the upper panel) are the predicted  
$t_e$(S {{\sc iii}}) -- $t_e$(O {{\sc iii}}) relations
for H {{\sc ii}} region models from \citet{Iz06}. 
 The dotted and thin solid black lines are the curves from \citet{Garnett1992}
and  \citet{PerezDiaz_05}, respectively. 
Regression lines (solid lines) and 1$\sigma$ alternatives (dashed lines) 
for all our data plus data from 
\citet{Garcia_Esteb2007} are shown by purple lines.   
 In the lower panel the error bars are shown for the VLT, SDSS, and
\citet{Garcia_Esteb2007} data.
Additionally, regression lines for VLT+SDSS-only data are shown by 
black lines.
\vspace{0.1cm}
\hspace{4.9cm} (A color version of this figure is available in the online journal.)
 }
\label{fig5}
\end{figure}

\setcounter{figure}{5}

\begin{figure}[t]
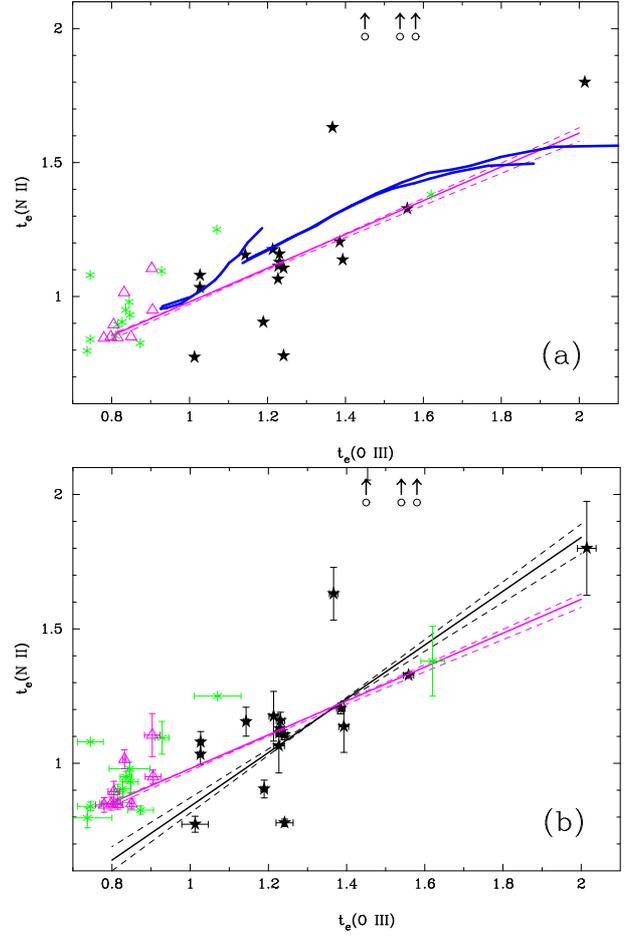

\hspace*{0.0cm}\psfig{figure=16291f6_1.ps,angle=-90,width=8.cm,clip=}
\hspace*{0.0cm}\psfig{figure=16291f6_2.ps,angle=-90,width=8.cm,clip=}
\caption{Comparison of electron temperatures $t_e$(O {{\sc iii}}) 
($t_e$ = 10$^{-4}$ $T_e$)
obtained from [O {{\sc iii}}]$\lambda$4363/($\lambda$4959 $+$ $\lambda$5007)
and $t_e$(N {{\sc ii}}) obtained from 
[N {{\sc ii}}]$\lambda$5755/$\lambda$6583 emission-line ratios.
VLT/UVES and FORS data are shown by stars, except for two objects,
which are shown as open circles (see text). 
  Green asterisks and open purple triangles provide the data
from \citet{Esteban2009} and \citet{Garcia_Esteb2007}, respectively.
 The thick blue lines show the predicted
$t_e$(O {{\sc ii}}) -- $t_e$(O {{\sc iii}}) relation
for H {{\sc ii}} region models from \citet{Iz06}. 
 Regression lines (solid lines) and 1$\sigma$ alternatives (dashed lines) 
for all data are shown by purple lines.
  The lower panel shows the error bars. 
 Additionally, regression lines for our VLT-only data are shown by 
black lines.
\vspace{0.1cm}
\hspace{10.5cm} (A color version of this figure is available in the online journal.)
}
\label{fig6}
\end{figure}

\setcounter{figure}{6}

\begin{figure}[t]
\hspace*{0.0cm}\psfig{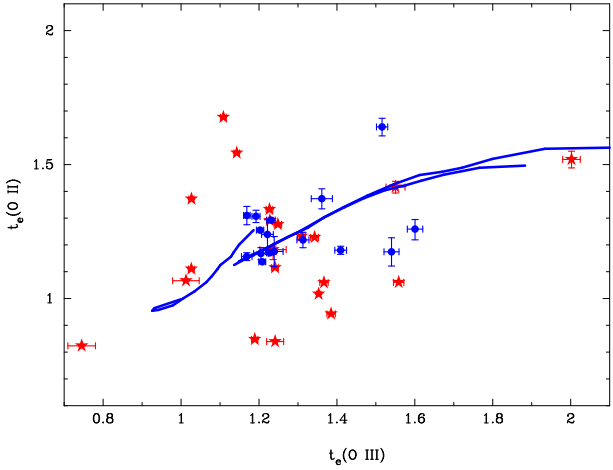}
\caption{Comparison of electron temperatures $t_e$(O {{\sc iii}}) 
($t_e$ = 10$^{-4}$ $T_e$)
derived from [O {{\sc iii}}]$\lambda$4363/($\lambda$4959 $+$ $\lambda$5007)
and $t_e$(O {{\sc ii}}) obtained from 
[O {{\sc ii}}]$\lambda$3727/($\lambda$7320 $+$ $\lambda$7330) 
emission-line ratios.
VLT/UVES and FORS data are shown by red stars and blue filled circles, 
respectively.
The thick blue lines represent the $t_e$(O {{\sc ii}}) -- $t_e$(O {{\sc iii}})
relations from \citet{Iz06}.
\vspace{0.1cm}
\hspace{10.5cm} (A color version of this figure is available in the online journal.)
}
\label{fig7}
\end{figure}

\setcounter{figure}{7}

\begin{figure}[t]
\hspace*{0.0cm}\psfig{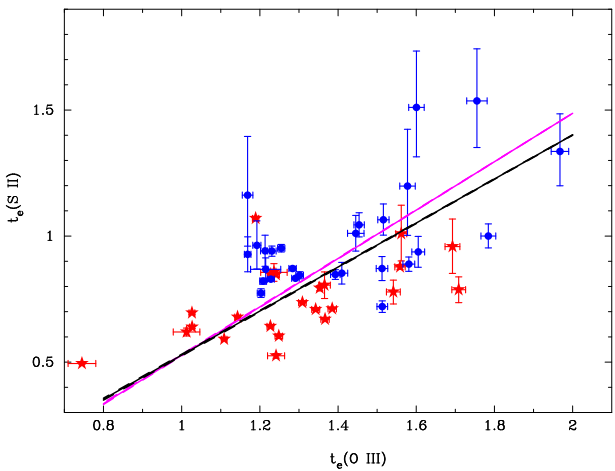}
\caption{Comparison of electron temperatures $t_e$(O {{\sc iii}}) 
($t_e$ = 10$^{-4}$ $T_e$)
obtained from [O {{\sc iii}}]$\lambda$4363/($\lambda$4959 $+$ $\lambda$5007)
and $t_e$(S {{\sc ii}}) obtained from 
[S {{\sc ii}}]$\lambda$4068/($\lambda$6717 $+$ $\lambda$6730) 
emission-line ratios.
Symbols are the same as in Fig~\ref{fig7}.
 Regression lines for all data are represented by purple lines.
 Regression lines for the UVES-only data are shown by black lines.
\vspace{0.1cm}
\hspace{1.5cm} (A color version of this figure is available in the online journal.)
}
\label{fig8}
\end{figure}

\section{Physical conditions in the H {\sc ii} regions}

 Physical conditions and element abundances in the selected H {\sc ii} 
regions were derived with the atomic data from the references listed in 
\citet{Stas2005}.

\subsection{Electron temperatures\label{te}}

 Nineteen out of the thirty one H {{\sc ii}} regions observed with VLT/UVES  (Table~\ref{tab2}) 
have detectable [S {{\sc iii}}]$\lambda$6312 and [S {{\sc iii}}]
$\lambda$9069 emission lines in their spectra. 
 This gives us the opportunity to determine the electron
temperature $T_e$(S {{\sc iii}}) directly from the spectra. 
  In Fig.~\ref{fig5} we compare the electron temperatures $t_e$(O {{\sc iii}}) and 
$t_e$(S {{\sc iii}}) ($t_e$ = 10$^{-4}$ $T_e$) obtained from 
[O {{\sc iii}}]$\lambda$4363/($\lambda$4959+$\lambda$5007) and 
[S {{\sc iii}}]$\lambda$6312/$\lambda$9069, respectively. 
  The H {{\sc ii}} region data from VLT/UVES, including
those for SBS 0335--052E from \citet{IGFP2009},  are shown by stars. 
 An exception are the three galaxies, which are represented 
by open circles.
One of them  (bottom left corner in both panels) belongs 
to the galaxy He 2--10E  (UVES, ESO program 081.C-0113(A)). 
Its spectrum is noisy with a very weak
and uncertain [O {{\sc iii}}]$\lambda$4363 line intensity.
 The other two, in the upper right corner, belong to Tol 1214--277
(UVES, ESO program 69.D-0174(A))  
and UM 133H (UVES, ESO program 68.B-0310(A)). At the redshift of 
Tol 1214--277 two absorption night-sky 
lines ($\lambda$9300.6 and $\lambda$9303.9) 
coincide with the brightest part of the 
 profile of the [S {{\sc iii}}]$\lambda$9069 emission line, 
reducing its flux by 
$\sim$20\% so that $t_e$(S {{\sc iii}}) is overestimated 
(the point is actually outside the figure, which is indicated 
by an arrow).
In UM133, the [S {{\sc iii}}]$\lambda$9069 emission line is affected by the night sky line
at  $\lambda$9118. 
 All our VLT/UVES and FORS data shown in Figs.~\ref{fig5}--\ref{fig11}
are presented in Table~\ref{tab10}.

  Additionally, we plot as red dots the data from \citet{Iz06} for the SDSS DR3.
 Out of more than $\sim$300 emission-line galaxies 
selected from the SDSS DR3 because they have a detectable 
[O {{\sc iii}}]$\lambda$4363 line, we 
show here only the 27 galaxies for which  $F$(H$\beta$) is 
  higher than 2$\times$10$^{-14}$ erg s$^{-1}$ cm$^{-2}$  
and for which the errors in the [O {\sc iii}]$\lambda$4363 fluxes are lower 
than 25\%. 
 For comparison we also show
data from  \citet{Kehrig06} (open blue 
rombs), \citet{PerezDiaz_03} (blue asterisks) and \citet{Garcia_Esteb2007} 
(purple triangles).
  The data by \citet{PerezDiaz_03} and \citet{Kehrig06}
were obtained from a spectroscopic analysis of H {\sc ii} regions in  
emission-line galaxies with ongoing star formation and the data by 
\citet{Garcia_Esteb2007}
were derived for  H {\sc ii} regions in our Galaxy.

The dashed blue line in the top panel connects points of equal 
temperatures. 
  The thick lines show the predicted 
$t_e$(S {{\sc iii}}) -- $t_e$(O {{\sc iii}}) relation
for H {{\sc ii}} region sequences of photoionization models with low, 
intermediate, and high metallicities 
(12+logO/H = 7.2, 7.6 and 8.2)
from \citet{Iz06}. 
 The dotted and solid black lines display the model prediction 
from \citet{Garnett1992}
and \citet{PerezDiaz_05}, respectively.   
  The regression line including our data, the SDSS data, and the data from 
\citet{Garcia_Esteb2007} are shown by purple lines. Its equation is 
\begin{eqnarray}
{t}_{e}({\rm SIII})=(-0.1183\pm0.0321)+(1.0590\pm0.0281){\times}
{t}_{e}({\rm OIII}).
\label{tSIII}
\end{eqnarray}

The regression was computed using the maximum likelihood method 
by \citet{Press1992} that includes error bars on both axes.
   In the lower panel the error bars concerning the VLT, SDSS,
\citet{Garcia_Esteb2007} data and additional regression lines for  
VLT and SDSS-only data are shown by black lines. 

Clearly the data presented here define a much better empirical relation between 
$t_e$(S {{\sc iii}}) and $t_e$(O {{\sc iii}}) than previous work on extragalactic 
H {{\sc ii}} regions. 
 We note that  the VLT/UVES data follow the same distribution 
as the best data from SDSS DR3.
Unfortunately, for high electron temperatures there are only few data with 
reliable temperature determinations. 

In metal-poor galaxies, [N {{\sc ii}}]$\lambda$5755 is extremely weak. 
Only in  19 out of 121 VLT spectra  could it be measured. 
To enlarge the range of temperatures we also added VLT/FORS data 
for the hot H {\sc ii} region in J0519+0007 from \citet{G2009}.
   In Fig.~\ref{fig6} we plot  $t_e$(N {{\sc ii}}) obtained from 
[N {{\sc ii}}]$\lambda$5755/$\lambda$6583 as 
a function of $t_e$(O {{\sc iii}}).  
 Data from our VLT sample are represented by stars. 
  Tol 2240-384 (FORS medium and low resolution) 
and Tol 2138-405 No.3 (FORS medium resolution)  are denoted by open 
circles with arrows, indicating that they are outside the plot
owing to large $t_e$(N {{\sc ii}}).
 Green asterisks display the data from \citet{Esteban2009}.
By open purple triangles we show the data 
from \citet{Garcia_Esteb2007}.

 The thick blue lines in the top panel of Fig.~\ref{fig6} represent the 
$t_e$(O {{\sc ii}}) -- $t_e$(O {{\sc iii}}) 
relation from the same  \citet{Iz06} models as those considered 
in Fig.~\ref{fig5}. 
 It roughly follows the observational points, which 
show a substantial scatter. 
However, at the lowest temperatures considered, there seems 
to be an offset between 
the blue lines and  our observational  points. 

The regression line for all data (upper panel) is given by the equation:
\begin{eqnarray}
{t}_{e}({\rm N II})=(0.3501\pm0.0158)+(0.6294\pm0.0138){\times}
{t}_{e}({\rm O III}).
\label{tNII}
\end{eqnarray}

  The lower panel shows the error bars. 
 Additionally, regression lines for our VLT-only data are shown by 
black lines.
  The electron temperature derived from 
[S {{\sc iii}}]$\lambda$6312/$\lambda$9069 or 
[N {{\sc ii}}]$\lambda$5755/$\lambda$6583  
allowed us to obtain
an empirical relation between those temperatures and the one derived from 
[O {{\sc iii}}]$\lambda$4363/$\lambda$(4959+5007)
for the first time for metal-poor galaxies.

In 21 UVES and 16 FORS spectra [O {{\sc ii}}]$\lambda$3727 and 
$\lambda$$\lambda$7320,7330 lines have been measured. 
The [S {{\sc ii}}]$\lambda$4068
and $\lambda$$\lambda$6717,6730 lines are detected and measured in 24 UVES and 
27 FORS spectra. These allow us to derive the electron 
temperatures $t_e$(O {{\sc ii}})
and $t_e$(S {{\sc ii}}) by direct methods. 
  The results are shown in Figs.~\ref{fig7}
and ~\ref{fig8}, where we compare those values with $t_e$(O {{\sc iii}}). 
 The UVES data 
are shown by red stars. Blue filled circles correspond to FORS data.
The thick blue lines in Fig.~\ref{fig7} represent the $t_e$(O {{\sc ii}}) --
$t_e$(O {{\sc iii}}) relations from \citet{Iz06}.
 Despite the large spread of points in Fig.~\ref{fig7} the VLT/UVES and FORS 
data follow the predicted $t_e$(O {{\sc ii}}) --
$t_e$(O {{\sc iii}}) relations \citep{Iz06}.
  The values of $t_e$(S {{\sc ii}}) in Fig.~\ref{fig8} correlate  well
with $t_e$(O {{\sc iii}}), but are significantly below $t_e$(O {{\sc ii}})
 in Fig.~\ref{fig7}.
Regression lines for all data in Fig.~\ref{fig8} are shown by purple lines
 and for the UVES-only data by black lines.
The regression line for all the data displayed in Fig.~\ref{fig8} 
is given by the equation:
\begin{eqnarray}
{t}_{e}({\rm S II})=(-0.4323\pm0.0029)+(0.9596\pm0.0014){\times}
{t}_{e}({\rm O III}).
\label{tSii}
\end{eqnarray}

\subsection{The electron number densities\label{ne}} 

\setcounter{figure}{8}

\begin{figure}[t]
\hspace*{0.0cm}\psfig{figure=16291f9_1.ps,angle=-90,width=8.cm,clip=}
\hspace*{0.0cm}\psfig{figure=16291f9_2.ps,angle=-90,width=8.cm,clip=}
\caption{$N_e$(O {{\sc ii}}) versus $N_e$(S {{\sc ii}}). 
Black stars: VLT-UVES; filled blue circles: FORS observations (medium resolution).
Purple triangles: data from \citet{Garcia_Esteb2007}, green asterisks: data from 
\citet{Esteban2009}. 
 The dashed blue line connects points of equal densities.
 The lower panel shows the error bars and regression lines
and their 1$\sigma$ alternatives 
for all data (purple lines) and for the VLT-only data (black lines).
\vspace{0.1cm}
\hspace{7.5cm} (A color version of this figure is available in the online journal.)
}
\label{fig9}
\end{figure}

\setcounter{figure}{9}

\begin{figure}[t]
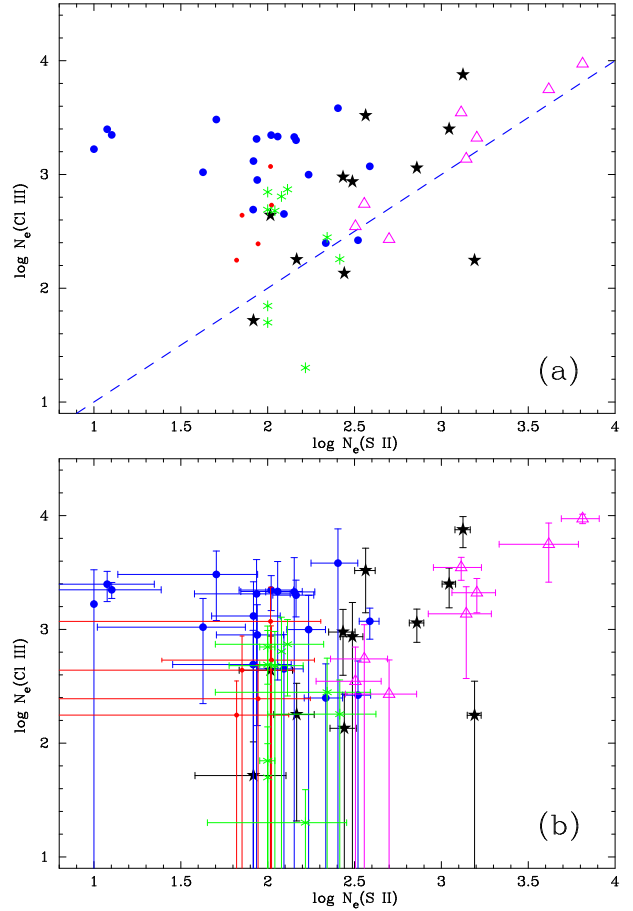

\hspace*{0.0cm}\psfig{figure=16291f10_1.ps,angle=-90,width=8.cm,clip=}
\hspace*{0.0cm}\psfig{figure=16291f10_2.ps,angle=-90,width=8.cm,clip=}
\caption{$N_e$(Cl {{\sc iii}}) versus  $N_e$(S {{\sc ii}}).
UVES, FORS 
and SDSS DR3 data are represented by
black stars, filled blue circles, 
and red dots, respectively. 
Purple triangles: data from \citet{Garcia_Esteb2007}, green asterisks: data from 
\citet{Esteban2009}.
The dashed blue line connects points of equal densities.
 The lower panel shows the error bars.
\vspace{0.1cm}
\hspace{0.0cm} (A color version of this figure is available in the online journal.)
}
\label{fig10}
\end{figure}

For all galaxies the electron number densities 
$N_e$(S {\sc ii}) were obtained from the [S {\sc ii}] 
$\lambda$6717/$\lambda$6731 emission line ratio, 
except for regions where these
emission lines are too weak 
to allow density determinations.
  For abundance determinations
in those regions we adopt $N_e$ = 10 cm$^{-3}$.
 The value of the electron number densities 
does not significantly affect
the derived abundances
in the low-density limit, which holds for the bulk of the H {\sc ii} regions
considered here. 
 The electron number densities $N_e$(S {\sc ii}) 
 derived from the [S {\sc ii}] 
$\lambda$6717/$\lambda$6731 emission line ratio
are given in Tables~\ref {tab6}, ~\ref{tab7} and ~\ref{tab8}.

The high signal-to-noise ratio and high spectral resolution of the 
spectra in the VLT sample 
allow us to measure the fluxes of the weak emission lines  
[Ar {{\sc iv}}]$\lambda$4711, 4740 and [Cl {{\sc iii}}]$\lambda$5517, 5537
and to separate and measure the fluxes of the 
[O {\sc ii}]$\lambda$3726,3729 lines to determine the electron number density.
  In many UVES spectra 
[O {\sc ii}]$\lambda$3726 and [O {\sc ii}]$\lambda$3729 are seen as 
separate lines (for instance in
UM 254, Fig.~\ref{fig2}). 
 However, for H {{\sc ii}} regions with broad 
components in their emission lines
these two lines are poorly separated. In those cases the IRAF 
routine SPLOT was used for deblending. 
 For all FORS medium resolution data the [O {\sc ii}]
$\lambda$3726+3729 lines were deblended by this IRAF software routine, 
adopting preliminary  estimated widths of each of these emission lines.

 The top panel of  Fig.~\ref{fig9} shows $N_e$(O {{\sc ii}}) versus  
$N_e$(S {{\sc ii}}). The data from UVES observations are shown by black stars, 
whereas medium-resolution
FORS data are denoted by filled blue circles.
 We did not include SDSS data because of the very large errors 
in $N_e$(O {{\sc ii}}) determinations.
Purple triangles represent data from \citet{Garcia_Esteb2007}, 
green asterisks data from \citet{Esteban2009}. 
 The dashed blue line connects points of equal densities.
The bottom panel of  Fig.~\ref{fig9} shows the same data as the 
upper panel, but with error bars and regression lines. 

The regression line for all data shown in Fig.~\ref{fig9} 
(lower panel, purple line) is presented by the equation
\begin{eqnarray}
{\rm log}{N}_{e}({\rm O II})=&(0.3413\pm1.2484)+(0.8782\pm0.4798) \nonumber\\
                             &{\times}{\rm log}{N}_{e}({\rm S II}).
\label{neOII}
\end{eqnarray}

The top panel of  Fig.~\ref{fig10} shows
$N_e$(Cl {{\sc iii}}) versus  $N_e$(S {{\sc ii}}).
UVES, FORS and SDSS DR3 data are represented by
black stars, filled blue circles and red dots, respectively. 
Purple triangles are the data from \citet{Garcia_Esteb2007}, 
green asterisks are the data from \citet{Esteban2009}.
 The dashed blue line displays locations with equal densities.
The error bars are shown in the lower panel.
Only 14 SDSS galaxies could be used to derive $N_e$(Cl {{\sc iii}}) due to our 
requirement that  the
error of the [Cl {{\sc iii}}]$\lambda$5517 emission line flux is less than 25\%. 
 Five of them are shown 
by red dots. 
 For the remaining nine, the SDSS data allow only the derivation of a 
lower limit of 10\,cm$^{-3}$ (they  are not shown in the Figure).

The top panel of  Fig.~\ref{fig11} shows
$N_e$(Ar {{\sc iv}}) versus  $N_e$(S {{\sc ii}}).
The FORS data are represented by filled blue circles.
Purple triangles are the data from \citet{Garcia_Esteb2007}, green asterisks 
are the data from \citet{Esteban2009}.
The dashed blue line is the place of equal densities.
 The lower panel shows the error bars and regression lines for all data.
  For   $N_e$(Ar {{\sc iv}}), we have a proper estimate for only  
two H {\sc ii} 
regions and a lower limit of 10\,cm$^{-3}$ for  
50 H {\sc ii} regions (not shown in the figure).

  For all the objects $N_e$(O {{\sc ii}}) 
 is very similar to $N_e$(S {{\sc ii}})
in the whole density range. Note that  H {{\sc ii}} regions with  $N_e$(S {{\sc ii}}) 
larger than 200\,cm$^{-3}$ represent only $\sim$10\% 
of the entire VLT sample (Tables~\ref {tab5} -- ~\ref{tab7}). 
On the other hand, $N_e$(Cl {{\sc iii}}) and $N_e$(Ar {{\sc iv}})
are systematically higher than $N_e$(S {{\sc ii}}). 
This has no consequence on our determination of heavy element 
abundances below, but it will be important for considering the 
derivation of the pregalactic helium abundance.

\section{Element abundances\label{chem}}

\subsection{Derivation of the abundances\label{chemcomp}}

 The above analysis shows that the relations $t_e$(S {\sc iii}) vs.
$t_e$(O {\sc iii}), $t_e$(N {\sc ii}) and $t_e$(O {\sc ii}) 
vs. $t_e$(O {\sc iii})
derived from observations follow the same trends as the models by 
\citet{Iz06}. Because the observed values are affected by large uncertainties 
in some cases and because many spectra lack the necessary data 
for the determination of $t_e$(S {\sc iii}) and  
$t_e$(N {\sc ii}), we prefer to use the same prescriptions 
as in Izotov et al. (2006).

  To derive the abundances of O$^{2+}$, Ne$^{2+}$ and Ar$^{3+}$, we adopted
the temperature $T_e$(O {\sc iii}) directly derived from the 
[O {\sc iii}] $\lambda$4363/($\lambda$4959 + $\lambda$5007)
emission-line ratio.

  For O$^{+}$,  N$^{+}$, S$^{+}$, and Fe$^{2+}$, we took the value of 
$t_e$(O {\sc ii}) obtained from $t_e$(O {\sc iii}) using Eq. 14  in  \citet{Iz06} that is shown below:
\begin{eqnarray}
t_e({\rm O\ II})=-0.577+t_e({\rm O\ III})\times[2.065-0.498t_e({\rm O\ III})], \nonumber\\
t_e({\rm O\ II})=-0.744+t_e({\rm O\ III})\times[2.338-0.610t_e({\rm O\ III})], \nonumber\\
t_e({\rm O\ II})=2.967+t_e({\rm O\ III})\times[-4.797+2.827t_e({\rm O\ III})] \label{toii},
\end{eqnarray}
for 12+logO/H=7.2 and $t_e$(O {\sc iii}) $\geq$ 1.14,
12+logO/H=7.6 and $t_e$(O {\sc iii}) $\geq$ 1.14 and
12+logO/H=8.2 and $t_e$(O {\sc iii}) $\leq$ 1.18, respectively
 (see thick blue lines in Fig. \ref{fig6}a).

Similarly, to compute the  S$^{2+}$, Cl$^{2+}$, and Ar$^{2+}$ abundances 
we adopted the value of $t_e$(S {\sc iii}) obtained 
from $t_e$(O {\sc iii}) using Eq. 15 in  \citet{Iz06}: 
 
\begin{eqnarray}
t_e({\rm S\ III})=-1.085+t_e({\rm O\ III})\times[2.320-0.420t_e({\rm O\ III})], \nonumber\\
t_e({\rm S\ III})=-1.276+t_e({\rm O\ III})\times[2.645-0.546t_e({\rm O\ III})], \nonumber\\
t_e({\rm S\ III})=1.653+t_e({\rm O\ III})\times[-2.261+1.605t_e({\rm O\ III})] \label{tsiii},
\end{eqnarray}
for 12+logO/H=7.2 and $t_e$(O {\sc iii}) $\geq$ 1.14,
12+logO/H=7.6 and $t_e$(O {\sc iii}) $\geq$ 1.14 and
12+logO/H=8.2 and $t_e$(O {\sc iii}) $\leq$ 1.18, respectively
(see thick blue lines in Fig. \ref{fig5}a).

Eqs. \ref{toii} and \ref{tsiii} 
\citep[identical with Eqs. 14 and 15 in ][]{Iz06} are based on the 
H {\sc ii} region-photoionization models calculated for the given
restricted (though large) ranges of 
input parameters and with the stellar atmosphere 
models by \citet{S02} for the three values of the metallicity corresponding to
12+logO/H = 7.2, 7.6 and 8.2 (see thick blue lines in 
Figs. \ref{fig6}a and \ref{fig5}a). 
Therefore they hold for the restricted ranges of 12+logO/H and 
$t_e$(O {\sc iii}) and we did not extrapolate them outside these ranges.
The electron temperatures $t_e$(O {\sc ii}) and $t_e$(S {\sc iii}) 
depend on both the oxygen abundance 12+logO/H and the electron 
temperature $t_e$(O {\sc iii}). 
 To derive these we fixed $t_e$(O {\sc iii})
and iteratively obtained 12+logO/H and $t_e$(O {\sc ii}). First, the approximate
value of $t_e$(O {\sc ii}) was derived from the first expression of 
Eq. \ref{toii} if $t_e$(O {\sc iii}) $\geq$ 1.14 or from the third expression
of Eq. \ref{toii} if $t_e$(O {\sc iii}) $<$ 1.14. Then the oxygen abundance 
was derived. Given the new iterative value of 12+logO/H, the new value of 
$t_e$(O {\sc ii}) was obtained. This process was continued until the 
convergence of both 12+logO/H and $t_e$(O {\sc ii}) was achieved. 

After that $t_e$(S {\sc iii}) was derived. 
We used the linear interpolation of $t_e$(O {\sc ii}) and $t_e$(S {\sc iii}) 
between 12+logO/H = 7.2, 7.6 and 8.2 to the iterative value of 12+logO/H. 
The interpolation ranges for 12+logO/H are between 7.2 and 8.2 for the
$t_e$(O {\sc iii}) range between 1.14 and 1.18 and are between 
7.2 and 7.6 for $t_e$(O {\sc iii}) $>$ 1.18. 
If the electron temperature $t_e$(O {\sc iii}) $<$ 1.14, we adopted the
equations for 12+logO/H=8.2 to derive $t_e$(O {\sc ii}) and $t_e$(S {\sc iii})
 (see Fig. \ref{fig6}a).
  If the iterated oxygen abundance
was above the 12 + logO/H = 8.2, we used the third lines in Eq.~\ref{toii} 
and ~\ref{tsiii} 
for the determination of $t_e$(O {\sc ii}) and $t_e$(S {\sc iii}).
  If the iterated 12 + logO/H $<$ 7.2 we used the first lines in  
Eqs.~\ref{toii} and ~\ref{tsiii} for the determination of $t_e$(O {\sc ii}) 
and $t_e$(S {\sc iii}).

\setcounter{figure}{10}

\begin{figure}[t]
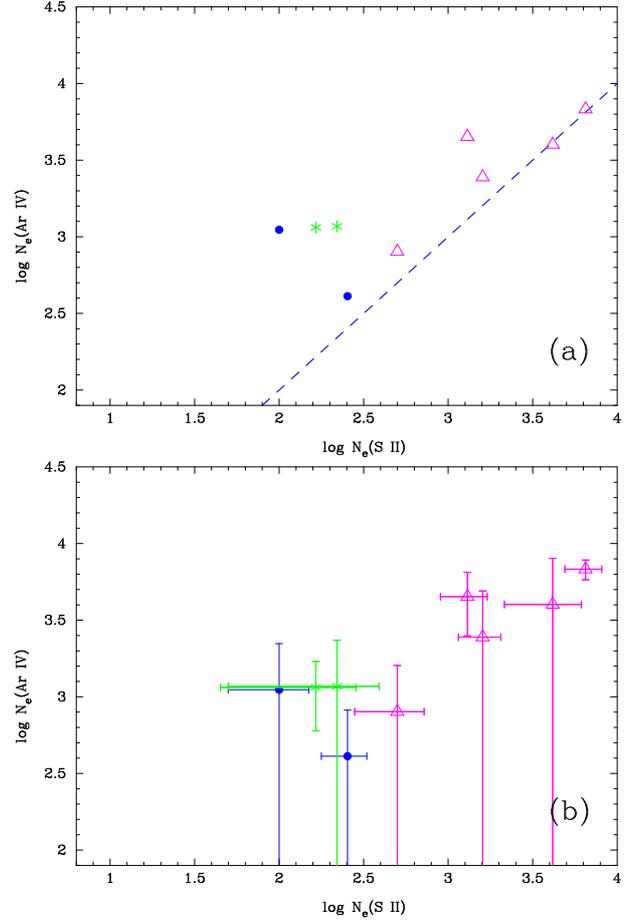

\hspace*{0.0cm}\psfig{figure=16291f11_1.ps,angle=-90,width=8.cm,clip=}
\hspace*{0.0cm}\psfig{figure=16291f11_2.ps,angle=-90,width=8.cm,clip=}
\caption{$N_e$(Ar {{\sc iv}}) versus  $N_e$(S {{\sc ii}}).
FORS data are represented by filled blue circles.
Purple triangles: data from \citet{Garcia_Esteb2007}, green asterisks: data from 
\citet{Esteban2009}.
  The dashed blue line connects points of equal densities.
  The lower panel shows the error bars.
\vspace{0.1cm}
\hspace{7.0cm} ~~~~~~~(A color version of this figure is available in the online journal.)
}
\label{fig11}
\end{figure}

\setcounter{figure}{11}

\begin{figure*}[t]
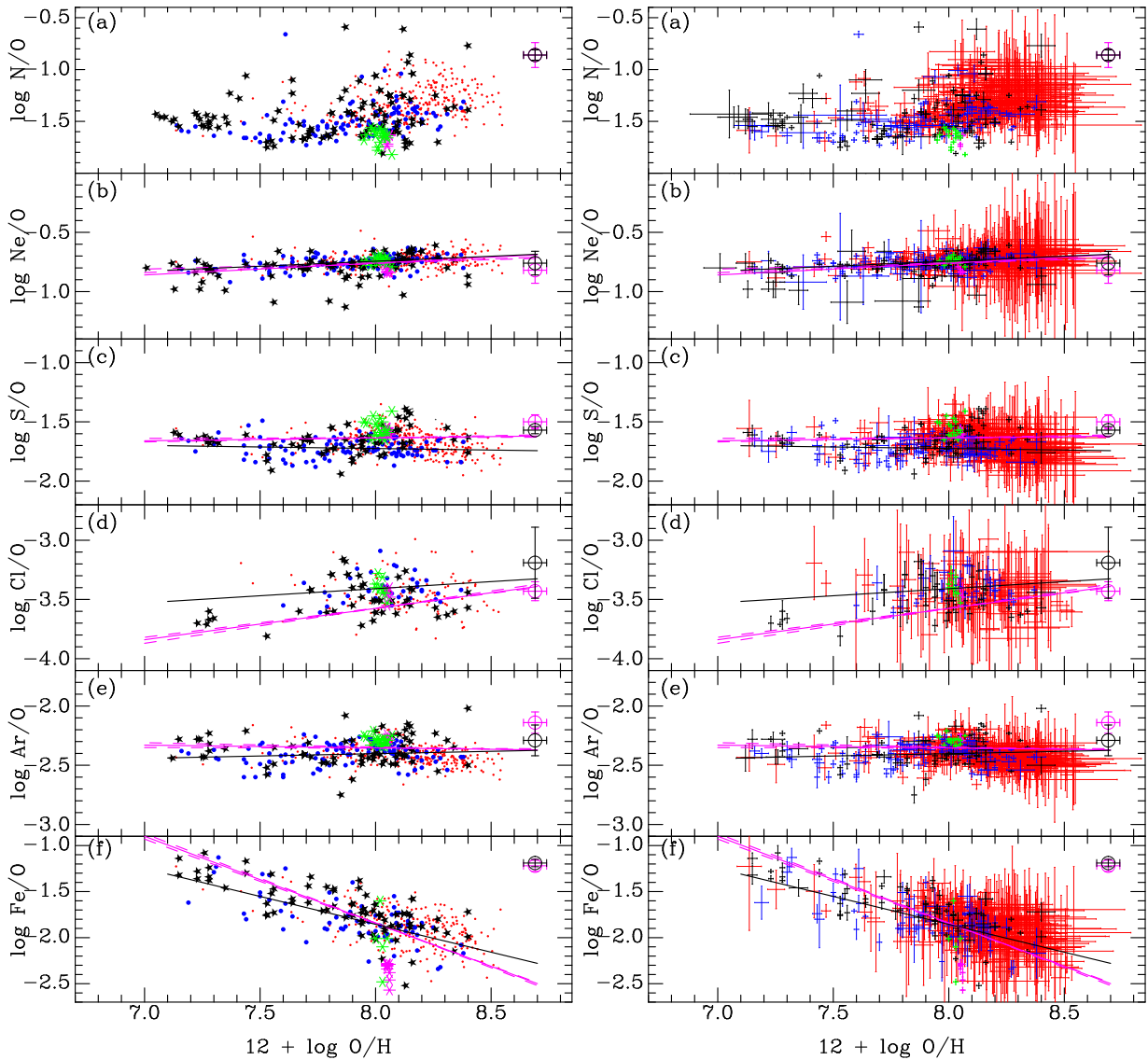

\hspace*{0.0cm}\psfig{figure=16291f12_1.ps,angle=0,width=8.cm,clip=}
\hspace*{0.0cm}\psfig{figure=16291f12_2.ps,angle=0,width=8.cm,clip=}
\caption{log N/O (a), log Ne/O (b), log S/O (c), log Cl/O (d),
log Ar/O (e) and log Fe/O (f) vs oxygen abundance
12 + log O/H for our sample of 121 spectra (large stars).
  Several knots in H {\sc ii} regions NGC 346 and NGC 456 in the SMC are
shown by green and purple large asterisks, respectively.
 Additionally we show  the galaxies from the
HeBCD sample of \citet{ISGT2004} and \citet{IT04a} for the 
primordial He abundance determination (small filled blue circles) and the 
sample of $\sim$300 emission-line galaxies from the SDSS DR3 \citep{Iz06}
(red dots).
The solar abundance ratios by \citet{Lodd2003} and by \citet{Asplund_09}
are indicated by the purple and black large open circles and the associated 
error bars.
The straight black lines are linear regressions for the 
HeBCD sample by \citet{Iz06} and the purple ones are for all 
(except the SDSS data). The right panel shows the error bars.
\vspace{0.05cm}
\hspace{12.cm} (A color version of this figure is available in the online journal.)
}
\label{fig12}
\end{figure*}

  The electron temperatures $T_e$(O {\sc iii}), $T_e$(O {\sc ii}), 
$T_e$(S {\sc iii}) and electron number density $N_e$(S {\sc ii})
are given in Table~\ref {tab5} (UVES observations) and 
Tables~\ref{tab6} and \ref{tab7} 
(low- and medium-resolution FORS observations).

All the ionic abundances were computed using the low-density limit 
of the emissivities.

 In all H {\sc ii} regions the abundances of O$^+$ and O$^{2+}$  
were obtained from the fluxes of the [O {\sc ii}] $\lambda$3727 and 
[O {\sc iii}]$\lambda$4959, 5007 lines,
respectively. 
  We added the small fraction of the undetected O$^{3+}$ ion to 
the oxygen abundance 
in high-excitation H {\sc ii} regions with O$^{+}$/(O$^{+}$+O$^{2+}$) $\le$ 0.1
if the He~{\sc ii} $\lambda$4686 emission line was detected in their spectra.
The total elemental abundances of the other elements were derived 
using the empirical ionization
correction factors ($ICF$s) of \citet{Iz06}) (their Eqs. 18--24).
 Similar to the electron temperatures, we used the linear 
interpolation of $ICF$s in the oxygen abundance range
12+logO/H = 7.2 -- 8.2 given the derived oxygen abundance for a particular
H {\sc ii} region. If the oxygen abundance of the H {\sc ii} region was greater
than 8.2, we adopted $ICF$s for 12+logO/H=8.2. Conversely, if the oxygen 
abundance of the H {\sc ii} region was less
than 7.2 we adopted $ICF$s for 12+logO/H=7.2.
 The ionic and total O, N, Ne, S, Cl, Ar, and Fe abundances derived from the
forbidden emission lines
are given in Tables~\ref {tab5}, ~\ref{tab6} and ~\ref{tab7} for the 
UVES and FORS observations.
The quoted errors in the element abundances account for the 
uncertainties in the ionization correction factors.

\subsection{Abundance patterns from collisionally excited lines (CELs)\label{chemCEL}} 

\setcounter{table}{8}

\begin{table*}
  \caption{Recombination lines: fluxes and ionic abundances \label{tab9}}
\begin{tabular}{lllrrlll} \hline
 Name &$F$(OII)$^{\rm a}$ &$F$(CII)$^{\rm a}$&$F$(H$\beta$)$^{\rm b}$ &&O$^{++}$/H$^{+}$$^{\rm c}$& O$^{++}$/H$^{+}$$^{\rm c}$&C$^{++}$/H$^{+}$$^{\rm c}$ \\ 
      &       &      &                &&(RL)  & (CEL) &(RL) \\ 
\hline \hline
NGC 346a No.1$^{\rm d}$&  7.77 $\pm$ 7.23 & $<$0.82           & 517.00$\pm$0.40  && 3.895$\pm$0.759& 0.888$\pm$0.021 &$<$0.165         \\
NGC 346a No.2$^{\rm d}$&  4.66 $\pm$ 0.67 & $<$0.58           & 381.30$\pm$0.34  && 3.124$\pm$0.120& 0.916$\pm$0.022 &$<$0.160         \\
NGC 346b No.1$^{\rm d}$&  5.49 $\pm$ 1.28 & $<$0.75           & 456.30$\pm$0.40  && 3.083$\pm$0.153& 0.859$\pm$0.021 &$<$0.172         \\
NGC 456a No.1$^{\rm d}$&  2.69 $\pm$ 0.65 &    1.43 $\pm$0.55 & 377.60$\pm$0.33  && 1.698$\pm$0.089& 0.703$\pm$0.017 & 0.392$\pm$0.151 \\
NGC 456a No.2$^{\rm d}$&  2.31 $\pm$ 0.77 &    2.22 $\pm$1.73 & 231.70$\pm$0.26  && 2.461$\pm$0.263& 0.857$\pm$0.022 & 0.981$\pm$0.762 \\
SBS 0335$-$052E$^{\rm e}$&~~~~~~~~...       &    3.20 $\pm$2.04 & 499.60$\pm$0.26  &&~~~~~~~~...     & 0.183$\pm$0.005 & 0.245$\pm$0.156 \\
He 2-10E$^{\rm e}$     & 22.07 $\pm$ 3.38 &$<$15.46           &3627.00$\pm$3.03  && 1.538$\pm$0.017& 1.471$\pm$0.297 &$<$0.171         \\ 
NGC 3125$^{\rm e}$     & 10.54 $\pm$ 2.18 &    9.07 $\pm$1.03 & 863.30$\pm$0.49  && 3.072$\pm$0.083& 1.507$\pm$0.034 & 0.426$\pm$0.049 \\
Mrk 1259$^{\rm e}$     & 41.30 $\pm$12.23 &$<$34.75           &1025.00$\pm$1.17  && 8.917$\pm$0.202& 0.568$\pm$0.064 &$<$1.336         \\
Mrk 1271$^{\rm e}$     & 10.78 $\pm$ 1.31 &    8.24 $\pm$2.82 &1368.00$\pm$0.89  && 1.957$\pm$0.048& 0.820$\pm$0.019 & 0.255$\pm$0.087 \\
Pox 4$^{\rm e}$        &  5.49 $\pm$ 1.90 &    3.30 $\pm$0.99 &1457.00$\pm$0.77  && 0.920$\pm$0.069& 1.031$\pm$0.024 & 0.096$\pm$0.029 \\
Tol 1214$-$277$^{\rm d}$ &  1.76 $\pm$ 0.49 & $<$0.22           & 700.80$\pm$0.51  && 0.521$\pm$0.027& 0.311$\pm$0.009 &$<$0.029         \\
Tol 1214$-$277$^{\rm e}$ & $<$3.41          & $<$0.75           & 700.80$\pm$0.52  &&$<$1.096        & 0.292$\pm$0.008 &$<$0.045         \\
J1253$-$0312$^{\rm e}$        & $<$3.29          &    2.34 $\pm$1.06 &1834.00$\pm$0.77  &&$<$0.359        & 0.869$\pm$0.020 & 0.054$\pm$0.024 \\
NGC5253 No.C1$^{\rm e}$   & 14.39 $\pm$ 2.35 &   11.95 $\pm$1.84 &2559.00$\pm$1.61  && 1.301$\pm$0.031& 1.195$\pm$0.027 & 0.196$\pm$0.030 \\
NGC5253 No.C2$^{\rm e}$   &  9.30 $\pm$ 1.92 &    9.88 $\pm$1.97 &1019.00$\pm$0.98  && 2.212$\pm$0.113& 1.487$\pm$0.035 & 0.384$\pm$0.077 \\
NGC5253 No.P1$^{\rm e}$   & 38.86 $\pm$12.83 &   28.99 $\pm$9.42 &1954.00$\pm$2.03  && 4.625$\pm$0.340& 0.729$\pm$0.011 & 0.626$\pm$0.203 \\
NGC5253 No.P2$^{\rm e}$   & 31.92 $\pm$ 7.78 &   23.38 $\pm$4.84 &3265.00$\pm$2.63  && 2.160$\pm$0.089& 1.194$\pm$0.027 & 0.303$\pm$0.063 \\
NGC 5408 No.1$^{\rm e}$   & 12.68 $\pm$ 1.72 &    7.20 $\pm$1.01 &1194.00$\pm$0.78  && 2.474$\pm$0.032& 0.939$\pm$0.022 & 0.255$\pm$0.036 \\
NGC 5408 No.2$^{\rm e}$   &  7.61 $\pm$ 0.99 &~~~~~~~~...        & 404.40$\pm$0.35  && 4.729$\pm$0.084& 1.073$\pm$0.026 &~~~~~~~~...        \\
Tol 1457$-$262$^{\rm e}$     & $<$3.25          &    3.75 $\pm$1.32 & 733.00$\pm$0.71  &&$<$1.218        & 1.312$\pm$0.030 & 0.213$\pm$0.075 \\
Tol1924$-$416 No.1$^{\rm e}$    & $<$6.77          &    3.26 $\pm$1.21 & 988.55$\pm$1.06  &&$<$1.568        & 0.784$\pm$0.019 & 0.140$\pm$0.052 \\
NGC 6822 V$^{\rm e}$    & 14.66 $\pm$ 1.76 &    6.94 $\pm$1.05 &1643.00$\pm$0.73  && 2.323$\pm$0.029& 1.265$\pm$0.027 & 0.173$\pm$0.026 \\
NGC 6822V No.1$^{\rm d}$&  1.36 $\pm$ 0.57 & $<$0.13           &  66.21$\pm$0.13  && 5.539$\pm$0.197& 1.192$\pm$0.033 &$<$0.212         \\
Tol 2138$-$405 No.1$^{\rm d}$&  0.53 $\pm$ 0.33 & $<$0.21           &  58.06$\pm$0.12  && 2.122$\pm$0.226& 0.855$\pm$0.022 &$<$0.382         \\
\hline \hline
\multicolumn{8}{l}{$^{\rm a}$ in units 10$^{-17}$ erg s$^{-1}$ cm$^{-2}$.}\\
\multicolumn{8}{l}{$^{\rm b}$ in units 10$^{-16}$ erg s$^{-1}$ cm$^{-2}$.}\\
\multicolumn{8}{l}{$^{\rm c}$ in units 10$^{-4}$.}\\
\multicolumn{8}{l}{$^{\rm d}$ medium resolution FORS data.}\\
\multicolumn{8}{l}{$^{\rm e}$ UVES data.}\\
\end{tabular}
\end{table*}

The derived oxygen abundances for our VLT sample are found
to cover the range of 12 + log O/H $\sim$ 7.2 -- 8.4.
  In Fig.~\ref{fig12} left panel, we show the abundance ratios versus 
metallicities for the 121 spectra from our VLT sample (large stars). 
 Several knots in H {\sc ii} regions NGC 346 and NGC 456 in the SMC are
shown by green and purple large asterisks, respectively.
 We also plot 
the data from the HeBCD sample collected by \citet{ISGT2004} and \citet{IT04a} 
(filled blue circles) and those from the DR3 of SDSS \citep{Iz06} (red dots).
 The HeBCD sample possesses more than 100 high-quality spectra of
low-metallicity galaxies for the primordial He abundance determination.
The error bars are shown in the right panel of the figure.

 In each panel, the solar abundance ratio by \citet{Lodd2003} 
is indicated by the purple large open circle along with the associated error 
bar, whereas the solar abundance ratio by \citet{Asplund_09} is indicated 
by the black large open circle.
 The straight black and purple lines in Fig.~\ref{fig12} are the linear 
regressions obtained for the HeBCD sample taken from the paper 
by \citet{Iz06} and for all data (121 VLT + 109 HeBCD spectra) respectively. 
 The SDSS data have 
larger errors (Fig.~\ref{fig12}, right panel) and therefore are not 
included in the regressions.

  The regression lines 
for all data (excluding SDSS) are given by the equations
\begin{eqnarray}
{\rm Ne}/{\rm O}=[(-1.2850\pm0.0472)+(0.0656\pm0.0059)]~ {\times}~{\rm X},
\label{Ne/O}
\end{eqnarray}
~~~~~~~(n = 249, $\chi$$^2$ = 1141),
\begin{eqnarray}
{\rm S}/{\rm O}=[(-1.8310\pm0.0532)+(0.0244\pm0.0066)]~{\times}~{\rm X},
\label{S/O}
\end{eqnarray}
~~~~~~~(n = 203, $\chi$$^2$ = 6126),
\begin{eqnarray}
{\rm Cl}/{\rm O}=[(-5.7188\pm0.0933)+(0.2680\pm0.0115)]~ {\times}~{\rm X},
\label{Cl/O}
\end{eqnarray}
~~~~~~~(n = 98, $\chi$$^2$ = 1701),
\begin{eqnarray}
{\rm Ar}/{\rm O}=[(-2.1739\pm0.0586)+(-0.0221\pm0.0071)]~ {\times}~{\rm X},
\label{Ar/O}
\end{eqnarray}
~~~~~~~(n = 196, $\chi$$^2$ = 5146),
\begin{eqnarray}
{\rm Fe}/{\rm O}=[(5.7180\pm0.0998)+(-0.9452\pm0.0121)]~{\times}~{\rm X},
\label{Fe/O}
\end{eqnarray}
~~~~~~~(n = 165, $\chi$$^2$ = 9032),\\
where X = 12 + log O/H.
 
Overall, we see in Fig.~\ref{fig12} a good overlap  
of our present VLT sample with the HeBCD and SDSS DR3 samples.
   The S/O, Cl/O and Ar/O abundance ratios are close to the distributions 
previously obtained by \citet{Iz06} and do not show any 
significant trend with oxygen abundance. Similar distributions for
S/O and Ar/O have been obtained by \citet{TIL95}, \citet{IT99} and
\citet{ZH06}.  
  We confirm the previous results
using a significantly larger data base that encompasses 230 high-quality data.
Some slight trends in S/O, Cl/O and Ar/O abundance ratios 
could imply a metallicity dependence 
of the $\alpha$-element production by massive stars. 
 Variations of the explosion energy of Type II supernovae 
with low metallicity might also play a role
\citep{K06}. However, \citet{Iz06}
pointed out that abundance determinations for some elements such as S, Cl, and
Ar could be uncertain because of the uncertainties in atomic data, 
e.g., in the rates of dielectronic recombination.

The ratio Ne/O (Fig.~\ref{fig12}{\bf b)} 
shows a slight increase with increasing 12 + log O/H. 
This is likely explained by a stronger depletion 
of oxygen onto dust grains in higher-metallicity galaxies.
 
 The most prominent trend was found by \citet{Iz06} for the Fe/O abundance 
ratio, which decreases with increasing O/H. While its value is nearly solar 
at the lowest 
metallicities, it drops by one order of magnitude at  12 + log O/H $=$ 8.5. 
 The interpretation of this trend put forward by  \citet{Iz06} is that 
it is due to depletion of Fe onto dust grains, which becomes 
more important at higher metallicities.
 
 The behavior of the N/O ratio is particularly interesting.  
\citet{IT99} and \citet{Iz06} have 
demonstrated that the dispersion in N/O in low-metallicity BCDs 
with 12 + log O/H $<$ 7.5 -- 7.6 is very low with a 
plateau value of log N/O $\sim$ --1.6. 
  However, in a larger sample (Fig.~\ref{fig12}{\bf a}) some galaxies 
have a higher N/O ratio at 12 + log O/H $<$ 7.6.
 Several points with highest N/O at 12 + log O/H $\sim$ 7.5 
(Fig.~\ref{fig12}{\bf a}) belong to galaxies with low $EW$(H$\beta$) (10 
to 70$\AA$). 
The notable exception is J0519+0007 with $EW$(H$\beta$) =240$\AA$. 
  This is in line with the finding of \citet{Iz06}
that the observed N/O increases as $EW$(H$\beta$) decreases.  
 They showed that this enhancement could be due to high-density 
nitrogen-rich ejecta from massive stars during W-R phases. The fact 
that no W-R features are detected in these HII regions is not a 
counterargument, because the W-R phase is expected to be short and the W-R 
features are weak in these low-metallicity systems.

  We also find that the lower limit of N/O in H {\sc ii} regions with 
lowest metallicity is higher than the above-mentioned plateau value 
of $-1.6$, implying that there is some increase 
in N/O with decreasing oxygen abundance at 12 + log O/H $<$ 7.5
(Fig.~\ref{fig12}{\bf a}).
If true, this tendency would agree with the prediction of
primary nitrogen being produced by low-metallicity rotating stars, as found by  
 \cite{MM02}.
Although these authors considered stellar models with a heavy element mass fraction
$Z$ = 10$^{-5}$, which is significantly lower than the values of 
$Z$ $\sim$ 0.0002 -- 0.0005 in
the lowest-metallicity H {\sc ii} regions analyzed by us,
it is possible that the interstellar medium in these galaxies
enriched by the first metal-free stars memorizes this.

  The two  H {\sc ii} regions in the SMC considered by us,
NGC 346 and NGC 456, have very low values  of 
log Fe/O, extending from --1.5 down to --2.6.
  It would be interesting to investigate why 
these  H {\sc ii} regions show such an extreme behavior. 
  The remaining elements (Ne, S, Cl, Ar) follow the same trends as 
in the other   
H {\sc ii} regions, with the exception of  N/O. NGC 346 and NGC 456 have the 
lowest  N/O ratio among all the H {\sc ii} regions in our sample.

\setcounter{figure}{12}

\begin{figure}[t]
\hspace*{0.0cm}\psfig{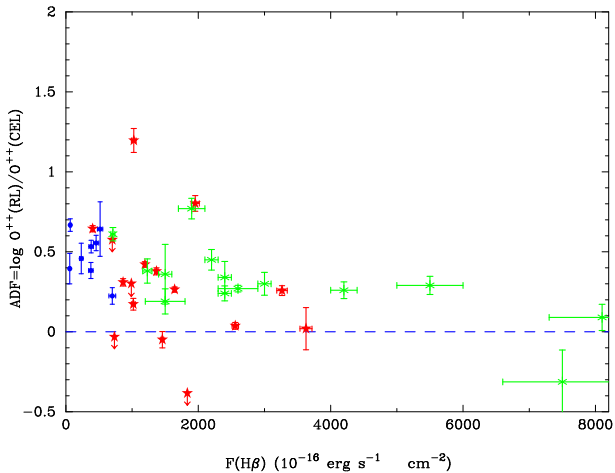}
\caption{Comparison of the abundance discrepancy factor
ADF = log O$^{++}$(RL)/O$^{++}$(CEL) with observed $F$(H$\beta$).
Data from UVES and FORS medium resolution observations are shown 
by red stars and filled blue circles.
Green asterisks are the data from \citet{Esteban2009}.
\vspace{0.1cm}
\hspace{1.5cm} (A color version of this figure is available in the online journal.)
}
\label{fig13}
\end{figure}

\setcounter{figure}{13}

\begin{figure}[t]
\hspace*{0.0cm}\psfig{figure=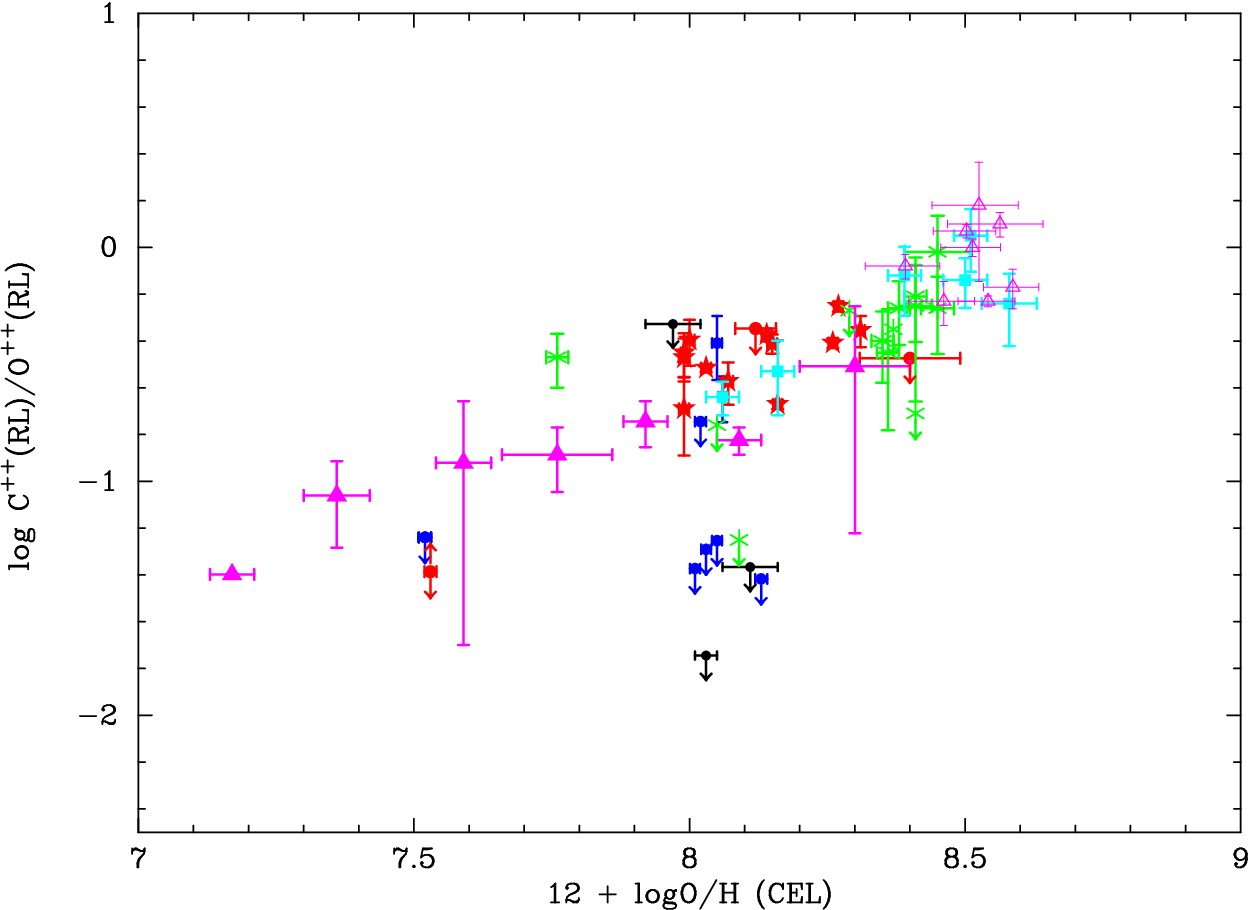,angle=-90,width=8.cm,clip=}
\caption{
Dependence of the ion abundance ratio log C$^{++}$(RL)/O$^{++}$(RL) 
on 12 + log O/H(CEL).
Our data are marked by the same symbols as in Fig.~\ref{fig13}.
Data from \citet{Garnett1995} and \citet{Garcia_Esteb2007} are shown 
by filled and open purple triangles, respectively.
Green asterisks are the data from \citet{Esteban2009}.
Data from \citet{Garnett1999} and \citet{KobulnSkill1998} are shown by 
light blue squares
and black circles, respectively.
\vspace{0.1cm}
\hspace{3.5cm} (A color version of this figure is available in the online journal.)
}
\label{fig14}
\end{figure}

\subsection{Ionic abundances from recombination lines \label{RLs}}

We detected and measured the fluxes of the recombination lines (RLs) 
O {{\sc ii}} $\lambda$4650 and C {{\sc ii}} $\lambda$4267 in 
eight medium-resolution FORS (out of 30) 
and 17 UVES (out of 31) H {\sc ii} region spectra. 
 We detected only two out of eight O {{\sc ii}} recombination 
lines of multiplet 1. The multiplet consists of eight lines:
$\lambda$4639, $\lambda$4642, $\lambda$4649, $\lambda$4651, $\lambda$4662, 
$\lambda$4674, $\lambda$4676, and $\lambda$4696. It is rarely possible to
measure all the lines of this multiplet, and frequently it is necessary 
to estimate the intensities of the unobserved or blended lines.
 Owing to the insufficient resolution of the FORS medium resolution spectra,
the O {{\sc ii}} $\lambda$4649 and O {{\sc ii}} $\lambda$4651 lines
are blended into the pair 4649+4651 named O {{\sc ii}} $\lambda$4650.
 In the UVES spectra we measured the O {{\sc ii}}$\lambda$4649 and 
$\lambda$4651 lines separately, which were then co-added and  
named O {{\sc ii}} $\lambda$4650. 
The intensities of O {{\sc ii}} $\lambda$4650 and 
C {{\sc ii}} $\lambda$4267 lines
were corrected for reddening using the extinction coefficient C(H$\beta$) 
from Tables~\ref{tab2} and ~\ref{tab4}.
Following  \citet{Esteban2009}  we calculated the abundance 
of O$^{++}$/H$^{+}$ from O {{\sc ii}} $\lambda$(4649+4651)
and  of C$^{++}$/H$^{+}$ from 
C {{\sc ii}} $\lambda$4267. 
 For the determination of the O$^{++}$ abundances 
we used the prescriptions by \citet{Peimbert2005}  
to calculate the
correction factor for the contribution of unmeasured lines of multiplet 1
for the non-LTE conditions. 
 The intensities of the lines of multiplet 1 weakly depend on the electron 
density. We used the values of  $N_e$(S {{\sc ii}}) to derive the 
correction factors. 
 If taking a higher value of the density, as indicated by 
$N_e$(Cl {{\sc iii}}) (Fig.~\ref{fig10}) and $N_e$(Ar {{\sc iv}}) 
(Fig.~\ref{fig11}), the 
RL abundances decrease by $\sim$5--15\%.
   The effective recombination coefficients were taken 
from \citet{Storey1994} for 
O {{\sc ii}} RLs and from \citet{Davey2000}  for C {{\sc ii}} RLs. 

  In Table~\ref{tab9} we present the observed fluxes of 
O {{\sc ii}} $\lambda$4650, C {{\sc ii}} $\lambda$4267 
and H$\beta$ emission lines,
the ionic abundances of O$^{++}$/H$^{+}$ obtained from RLs and CELs, and C$^{++}$/H$^{+}$ 
obtained from RLs using the electron temperature derived from  
[O {\sc iii}] $\lambda$4363/($\lambda$4959 + $\lambda$5007).

   O$^{++}$ is the only ion for which both strong CELs and 
measured RLs are present in the optical range.
 In Fig.~\ref{fig13} we show the dependence of the abundance discrepancy 
factor ADF$=$log O$^{++}$(RL)/O$^{++}$(CEL) on the observed $F$(H$\beta$). 
UVES data are shown by red stars, 
FORS medium-resolution data by filled blue circles.
 We determined, where possible, ADF and 
C$^{++}$(RL) and compared our results with those of other studies.
Green asterisks represent the data from \citet{Esteban2009}.
 There is no clear trend with $F$(H$\beta$) for 
$F$(H$\beta$) $>$ 10$^{-13}$ erg s$^{-1}$ cm$^{-2}$
and the error bars are large. 
Many UVES cases are compatible with no discrepancy at all. 
  However, half of the cases, 
especially the FORS and UVES data with 
$F$(H$\beta$) $<$ 10$^{-13}$ erg s$^{-1}$ cm$^{-2}$,
 indicate an ADF clearly larger than zero.  This is in 
line with what was found by Esteban et al. (2009)  for a set of extragalactic 
H {\sc ii} regions as well with what is summarized 
by  Garc\'ia-Rojas \& Esteban (2007) 
for galactic  H {\sc ii} regions.  We investigated the possible existence of 
a correlation between the ADF and any of the following 
parameters: $C$(H$\beta$),  
O$^{+}$/(O$^{+}$+O$^{+}$), O/H (CEL), FWHM(H$\beta$) and electron 
number density, 
hoping to find a clue for the origin of those ADFs. Like 
Garc\'ia-Rojas \& Esteban (2007), we found no evidence for any 
correlation whatsoever.

Whatever the reason for the abundance discrepancy may be, we can use the 
C$^{++}$(RL)/O$^{++}$(RL) ratio to achieve some insight into the behavior of 
the C/O ratio. Figure~\ref{fig14} shows the values of 
log C$^{++}$(RL)/O$^{++}$(RL) as a function of 12 + log O/H(CEL). 
The symbols are the same as in Fig.~\ref{fig13}.
  We supplement our data with those 
from \citet{Garnett1995}, where C$^{++}$/O$^{++}$ are derived from
the UV CELs and \citet{Garcia_Esteb2007}, which are shown 
by filled and open purple triangles, respectively.
Green asterisks are the data from \citet{Esteban2009}.
  The data from \citet{Garnett1999} and \citet{KobulnSkill1998} 
for the  C {\sc iii}]1909 line are shown by light blue squares
and black circles, respectively. We find that if C$^{++}$(RL)/O$^{++}$(RL) 
can be taken as a measure of C/O, the C/O ratio increases with O/H. 
 The only objects that do not follow this trend are 
the ones from our sample and from \citet{Esteban2009} and 
\citet{KobulnSkill1998}, for which only 1$\sigma$ upper limits to the
C {{\sc ii}} $\lambda$4267 line flux are estimated.
   The two leftmost points are for Tol 1214--277.
  One of these points is from the UVES spectrum, where only upper limits to the 
$F$(O {{\sc ii}}) and $F$(C {{\sc ii}}) are obtained, and the second point
is from the FORS medium-resolution spectrum.
Four points at 12 + log O/H $\sim$ 8.05 belong to NGC 6822 V No.1
and to three knots in the SMC H {\sc ii} 
region NGC 346. Note, that for NGC 346 
no night sky subtraction was made.

\section{Summary}

    We have presented an analysis of archival VLT/FORS1+UVES 
spectroscopic observations of a large sample of low-metallicity emission-line galaxies.
The whole sample, which contains some data from our previous 
papers (\citet{IGFP2009}, and \citet{G2009}), consists of  121 
spectra, out of which 83 are analyzed for the first time. 
For comparison, we also used data from SDSS DR3 studied by \citet{Iz06}
and 109 spectra from
the HeBCD sample observed with different telescopes and collected 
by \citet{IT04a} and \citet{ISGT2004} for the study of the primordial He 
abundance. 

Our main results are as follows:

1. The oxygen abundance   in the sample lies in the range 
12 + log O/H = 7.2 -- 8.4. 
The abundance ratios of the $\alpha$-elements to oxygen 
follow the trends found in our previous studies of 
low-metallicity emission-line galaxies. 
  In particular, the new data confirm with a larger sample 
of 230 H {\sc ii} regions the finding by \citet{Iz06} that
Ne/O increases with increasing oxygen abundance, which is interpreted as 
caused by a higher
depletion of oxygen in higher-metallicity galaxies. 
  The S/O, Cl/O and Ar/O abundance ratios are close to the distributions 
obtained previously by \citet{Iz06}.
 Slight trends seen in S/O, Cl/O and Ar/O abundance ratios 
could imply a metallicity dependence 
of the $\alpha$-element production by massive stars. 
 Variations of the explosion energy of Type II supernovae 
with metallicity might also play a role
\citep{K06}. However, as already pointed out by \citet{Iz06}, the abundance 
determinations of these elements may suffer from uncertainties 
owing to inaccurate ionization correction factors. 

2.  The Fe/O ratio shows an underabundance 
of iron relative to oxygen as compared to the solar value, 
which is particularly large for the high-metallicity galaxies. This again confirms 
the finding by  \citet{Iz06} and 
strengthens our interpretation that Fe is depleted onto dust grains
and that this effect depends on the metallicity.

3. There is a tendency for N/O to increase with decreasing oxygen abundance in
extremely low-metallicity galaxies with 12 + log O/H $<$ 7.5. 
This could be a sign of 
enhanced production of primary nitrogen by 
rapidly rotating metal-free stars \citep{MM02}.

4. The electron temperature derived from 
[S {{\sc iii}}]$\lambda$6312/$\lambda$9069 or 
[N {{\sc ii}}]$\lambda$5755/$\lambda$6583  
in a number of objects allowed us to obtain for the first 
time in metal-poor galaxies 
an empirical relation between those temperatures and the one derived from 
[O {{\sc iii}}]$\lambda$4363/$\lambda$(4959+5007).
 We also present the empirical relation between $t_e$ derived from 
[O {{\sc ii}}]$\lambda$3727/($\lambda$7320 $+$ $\lambda$7330) or 
[S {{\sc ii}}]$\lambda$4068/($\lambda$6717 $+$ $\lambda$6730) and 
[O {{\sc iii}}]$\lambda$4363/$\lambda$(4959+5007).

5. The electron number densities $N_e$(O {{\sc ii}}), $N_e$(Cl {{\sc iii}}),
$N_e$(Ar {{\sc iv}}) could be obtained for a number of objects in addition to 
$N_e$(S {{\sc ii}}). We find that $N_e$(O {{\sc ii}}) is very similar to  
$N_e$(S {{\sc ii}}), while $N_e$(Cl {{\sc iii}}) and $N_e$(Ar {{\sc iv}})
are systematically higher. This has potential implications when deriving the 
pregalactic helium abundance, since the He I lines are predominantly emitted in 
those higher density zones.  

6. In a number of objects, the abundances of C$^{++}$ and O$^{++}$ could be derived 
from recombination lines (RLs). We find that 
 O$^{++}$ abundances obtained from RLs tend to be  higher than 
those derived from collisionally excited lines (CELs), 
as found in previous studies of galactic and a few 
extragalactic H\,{\sc ii} regions. In the C$^{++}$/O$^{++}$ vs O/H diagram, 
most of the new points follow the relation obtained by previous observations 
of C/O increasing with O/H.

\begin{acknowledgements}
N. G. G. and Y. I. I., K. J. F. are grateful to the staff of the 
Max-Planck-Institute f\"ur  
Radioastronomie (Bonn) for their hospitality and acknowledge support 
through DFG grant No. FR 325/59-1. 
Y. I. I. thanks the Observatoire de Paris for hospitality and financial support.
P. P. has been supported by a Ciencia 2008 contract, funded by 
FCT/MCTES (Portugal) and POPH/FSE (EC), and by the Wenner-Gren Foundation.
\end{acknowledgements}

\setcounter{table}{1}

\begin{longtable}{lcccrcc}
\caption{Spectroscopic properties of H\,{\sc ii} regions.}  
\label{tab8}   \\
\hline \hline
\endfirsthead
\caption{\sl ---Continued.} \\
\hline \hline
\endhead
\endfoot
H\,{\sc ii} Region&$t_e$(O {\sc iii})$^{\rm a}$&12 + logO/H&EW(H$\alpha$)$^{\rm b}$&$F$(H$\beta$)$^{\rm c}$&$C$(H$\beta$) dex&100$\times$$F$([O {\sc iii}]4363)/$F$(H$\beta$)$^{\rm d}$ \\ \hline
UM 254$^{\rm e}$&1.07 $\pm$0.03 &8.16 $\pm$ 0.03&  402 $\pm$0  & 356.7$\pm$0.5 &0.130 $\pm$ 0.018 & 1.7 $\pm$ 0.1 \\
UM 283D$^{\rm e}$&1.24 $\pm$0.03 &8.07 $\pm$ 0.03&  314 $\pm$0  & 107.3$\pm$0.7 &0.225 $\pm$ 0.020 & 3.8 $\pm$ 0.3 \\
UM 283O$^{\rm e}$&1.18 $\pm$0.03 &8.06 $\pm$ 0.03&  301 $\pm$0  &  47.9$\pm$0.1 &0.290 $\pm$ 0.019 & 3.2 $\pm$ 0.2 \\
NGC 346 A No.1$^{\rm g}$&1.23 $\pm$0.01 &8.07 $\pm$ 0.01& 1157 $\pm$1  & 602.5$\pm$0.6 &0.060 $\pm$ 0.018 & 6.5 $\pm$ 0.1 \\
NGC 346 A No.2$^{\rm g}$&1.27 $\pm$0.01 &8.03 $\pm$ 0.01& 1330 $\pm$1  & 414.4$\pm$0.5 &0.105 $\pm$ 0.018 & 6.6 $\pm$ 0.1 \\
NGC 346 A No.3$^{\rm g}$&1.24 $\pm$0.01 &8.04 $\pm$ 0.01&  392 $\pm$0  & 339.8$\pm$0.6 &0.065 $\pm$ 0.018 & 6.0 $\pm$ 0.2 \\
NGC 346 B No.1$^{\rm g}$&1.24 $\pm$0.01 &8.05 $\pm$ 0.01& 1121 $\pm$1  & 470.5$\pm$0.8 &0.060 $\pm$ 0.019 & 5.8 $\pm$ 0.1 \\
\end{longtable}

\twocolumn



\Online

\setcounter{table}{2}
 \begin{longtable}{lrrrr}
\caption{Extinction-corrected emission line fluxes (UVES observations) \label{tab2}} \\
\hline \hline
\endfirsthead
\caption{\sl ---Continued.} \\
\hline \hline
\endhead
\endfoot
  &\multicolumn{4}{c}{\sc Galaxy} \\      
  &\multicolumn{4}{c}{\sc (ESO Program)}\\       \cline{2-5}
 Line&UM 254  &UM 283D &UM 283O & UM 133H \\
  &
(71.B-0055(A)) & (71.B-0055(A)) & (70.B-0717(A)) & (68.B-0310(A)) \\
\hline
3727 [O {\sc ii}]                 & 250.92 $\pm$   3.88 & 294.65 $\pm$   5.03 & 236.44 $\pm$   3.76 & 107.18 $\pm$   1.70 \\   
3750 H12                          &   3.00 $\pm$   0.05 &   3.58 $\pm$   0.87 &   5.29 $\pm$   0.67 &   3.95 $\pm$   0.29 \\
3771 H11                          &   3.20 $\pm$   0.19 &   4.78 $\pm$   0.49 &   6.20 $\pm$   0.43 &   5.08 $\pm$   0.29 \\
3798 H10                          &   5.55 $\pm$   0.29 &   6.11 $\pm$   0.53 &   8.14 $\pm$   0.39 &   5.54 $\pm$   0.30 \\
\end{longtable}

\setcounter{table}{3}
\begin{longtable}{lrrrr}
  \caption{Extinction-corrected emission line fluxes (low-resolution 
FORS observations) \label{tab3}} \\
\hline \hline
\endfirsthead
\caption{\sl ---Continued.} \\
\hline \hline
\endhead
\endfoot
  &\multicolumn{4}{c}{\sc Galaxy} \\      
  &\multicolumn{4}{c}{\sc (ESO Program)}\\       \cline{2-5}
  Line&
 NGC 346A No.1     &NGC 346A No.2     &
 NGC 346A No.3     &NGC 346B No.1  \\
 &
(69.C-0203(A)) & (69.C-0203(A)) & (69.C-0203(A)) & (69.C-0203(A)) \\
\hline   
3727 [O {\sc ii}]                 &  75.53 $\pm$   1.27 &  94.10 $\pm$   1.61 & 105.08 $\pm$   2.08 & 133.94 $\pm$   2.33 \\
3750 H12                          &   4.80 $\pm$   0.43 &   4.62 $\pm$   0.58 &     ...~~~~~~       &   6.11 $\pm$   1.15 \\
3771 H11                          &   5.45 $\pm$   0.37 &   5.30 $\pm$   0.44 &     ...~~~~~~       &   6.83 $\pm$   0.80 \\
3798 H10                          &   6.14 $\pm$   0.31 &   6.58 $\pm$   0.42 &     ...~~~~~~       &   7.35 $\pm$   0.70 \\
\end{longtable}

\setcounter{table}{4}
\begin{longtable}{lrrrr}
  \caption{Extinction-corrected emission line fluxes (medium-resolution 
FORS observations) \label{tab4}} \\
\hline \hline
\endfirsthead
\caption{\sl ---Continued.} \\
\hline \hline
\endhead
\endfoot
  &\multicolumn{4}{c}{\sc Galaxy} \\      
  &\multicolumn{4}{c}{\sc (ESO Program)}\\       \cline{2-5}
  Line&
 NGC 346A No.1       &NGC 346A No.2       &
 NGC 346A No.3       &NGC 346B No.1 \\
 &
(69.C-0203(A)) & (69.C-0203(A)) & (69.C-0203(A)) & (69.C-0203(A)) \\
\hline
3727 [O {\sc ii}]                 &  89.43 $\pm$   1.39 &  94.81 $\pm$   1.47 & 116.32 $\pm$   1.81 & 155.64 $\pm$   2.41 \\
3750 H12                          &   4.54 $\pm$   0.13 &   3.54 $\pm$   0.11 &   2.95 $\pm$   0.23 &   4.53 $\pm$   0.14 \\
3771 H11                          &   5.24 $\pm$   0.13 &   4.38 $\pm$   0.12 &   4.00 $\pm$   0.21 &   5.16 $\pm$   0.14 \\
3798 H10                          &   6.96 $\pm$   0.15 &   5.62 $\pm$   0.13 &   4.83 $\pm$   0.20 &   6.47 $\pm$   0.15 \\
\end{longtable}

\setcounter{table}{5}
 \begin{longtable}{lrrrr}
  \caption{Ionic and total heavy element abundances (UVES observations)
\label{tab5}} \\
\hline \hline
\endfirsthead
\caption{\sl ---Continued.} \\
\hline \hline
\endhead
\endfoot
  &\multicolumn{4}{c}{\sc Galaxy}\\       
  &\multicolumn{4}{c}{\sc (ESO Program)}\\       \cline{2-5}
  Property&
\multicolumn{1}{c}{UM 254   } & \multicolumn{1}{c}{UM 283D   } &
\multicolumn{1}{c}{UM 283O   } & \multicolumn{1}{c}{UM 133H} \\
  &
\multicolumn{1}{c}{(71.B-0055(A))} & \multicolumn{1}{c}{(71.B-0055(A))} &
\multicolumn{1}{c}{(70.B-0717(A))} & \multicolumn{1}{c}{(68.B-0310(A))} \\
\hline
  $T_{\rm e}$(O {\sc iii}) (K) &
$10666 \pm 260$ & $12360 \pm 338$ &
$11761 \pm 293$ & $15504 \pm 243$ \\

  $T_{\rm e}$(O {\sc ii}) (K) &
$10666 \pm 287$ & $12139 \pm 381$ &
$12082 \pm 328$ & $14146 \pm 287$ \\

  $T_{\rm e}$(S {\sc iii}) (K) &
$10673 \pm 215$ & $11591 \pm 281$ &
$11639 \pm 243$ & $15124 \pm 202$ \\

  $N_{\rm e}$(S {\sc ii}) (cm$^{-3}$) &
$ 10  \pm  10 $ & $ 15  \pm  15 $ &
$ 10  \pm  10 $ & $ 59  \pm  40 $  \\
\hline
\end{longtable}

\setcounter{table}{6}
\begin{longtable}{lrrrr}
  \caption{Ionic and total heavy element abundances (FORS low-resolution observations)
\label{tab6}} \\
\hline \hline
\endfirsthead
\caption{\sl ---Continued.} \\
\hline \hline
\endhead
\endfoot
  &\multicolumn{4}{c}{\sc Galaxy}\\       
  &\multicolumn{4}{c}{\sc (ESO Program)}\\       \cline{2-5}
  Property&
\multicolumn{1}{c}{NGC 346A No.1     } & \multicolumn{1}{c}{NGC 346A No.2     } &
\multicolumn{1}{c}{NGC 346A No.3     } & \multicolumn{1}{c}{NGC 346B No.1} \\
  &
\multicolumn{1}{c}{(69.C-0203(A))} & \multicolumn{1}{c}{(69.C-0203(A))} &
\multicolumn{1}{c}{(69.C-0203(A))} & \multicolumn{1}{c}{(69.C-0203(A))} \\
\hline
  $T_{\rm e}$(O {\sc iii}) (K) &
$12300  \pm  94 $ & $12685  \pm 105 $ &
$12370  \pm 149 $ & $12392  \pm 123 $  \\

  $T_{\rm e}$(O {\sc ii}) (K) &
$12088  \pm 106 $ & $12402  \pm 119 $ &
$12147  \pm 168 $ & $12166  \pm 138 $  \\

  $T_{\rm e}$(S {\sc iii}) (K) &
$11513  \pm  78 $ & $12007  \pm  87 $ &
$11604  \pm 124 $ & $11633  \pm 102 $  \\

  $N_{\rm e}$(S {\sc ii}) (cm$^{-3}$) &
$   10  \pm  10 $ & $   52  \pm  31 $ &
$   77  \pm  36 $ & $   36  \pm  29 $  \\
\hline
\end{longtable}

\setcounter{table}{7}
\begin{longtable}{lrrrr}
  \caption{Ionic and total heavy element abundances (FORS medium-resolution observations)
\label{tab7}} \\
\hline \hline
\endfirsthead
\caption{\sl ---Continued.} \\
\hline \hline
\endhead
\endfoot
  &\multicolumn{4}{c}{\sc Galaxy}\\       
  &\multicolumn{4}{c}{\sc (ESO Program)}\\       \cline{2-5}
  Property&
\multicolumn{1}{c}{NGC 346A No.1       } & \multicolumn{1}{c}{NGC 346A No.2       } &
\multicolumn{1}{c}{NGC 346A No.3       } & \multicolumn{1}{c}{NGC 346B No.1       } \\
  &
\multicolumn{1}{c}{(69.C-0203(A))} & \multicolumn{1}{c}{(69.C-0203(A))} &
\multicolumn{1}{c}{(69.C-0203(A))} & \multicolumn{1}{c}{(69.C-0203(A))} \\
\hline
  $T_{\rm e}$(O {\sc iii}) (K) &
$12955  \pm  97 $ & $12563  \pm  93 $ &
$12666  \pm 106 $ & $12594  \pm  94 $ \\
  $T_{\rm e}$(O {\sc ii}) (K) &
$12611  \pm 110 $ & $12305  \pm 105 $ &
$12387  \pm 120 $ & $12330  \pm  87 $ \\
  $T_{\rm e}$(S {\sc iii}) (K) &
$12343  \pm  80 $ & $11852  \pm  77 $ &
$11982  \pm  88 $ & $11892  \pm  78 $ \\
  $N_{\rm e}$(S {\sc ii}) (cm$^{-3}$) &
$   10  \pm  10 $ & $   42  \pm  32 $ &
$   51  \pm  37 $ & $   65  \pm  35 $ \\
\hline
\end{longtable}

\newpage
\setcounter{table}{9}
\begin{longtable}{lrrrrr}
\caption{Electron temperatures and electron number densities derived from emission line flux ratios.}  
\label{tab10}   \\
\hline
\endfirsthead
\caption{\sl ---Continued.} \\
\hline
\endhead
\endfoot
Name                           & $t_e$(O {{\sc iii}})$^{\rm a}$ & $t_e$(S {{\sc iii}})$^{\rm a}$ & $t_e$(N {{\sc ii}})$^{\rm a}$ & $t_e$(O {{\sc ii}})$^{\rm a}$   & $t_e$(S {{\sc ii}})$^{\rm a}$   \\ \hline
UM 283D$^{\rm e}$              & 1.2360 $\pm$ 0.0338            & 1.1116 $\pm$ 0.0603            &\multicolumn{1}{c}{...}        & 1.1827 $\pm$ 0.0251   & 0.8560 $\pm$ 0.0350   \\   
NGC 346 D No.1$^{\rm g}$       & 1.3253 $\pm$ 0.0126            &\multicolumn{1}{c}{...}         &\multicolumn{1}{c}{...}        &\multicolumn{1}{c}{...}& 0.8326 $\pm$ 0.0041   \\
NGC 346 D No.2$^{\rm g}$       & 1.3332 $\pm$ 0.0123            &\multicolumn{1}{c}{...}         &\multicolumn{1}{c}{...}        &\multicolumn{1}{c}{...}& 0.9519 $\pm$ 0.0148   \\
NGC 346 D No.3$^{\rm g}$       & 1.2741 $\pm$ 0.0116            &\multicolumn{1}{c}{...}         &\multicolumn{1}{c}{...}        &\multicolumn{1}{c}{...}& 0.8714 $\pm$ 0.0085   \\
NGC 346 D No.4$^{\rm g}$       & 1.3036 $\pm$ 0.0121            &\multicolumn{1}{c}{...}         &\multicolumn{1}{c}{...}        &\multicolumn{1}{c}{...}& 0.8445 $\pm$ 0.0141   \\
NGC 456 A No.1$^{\rm g}$       & 1.2308 $\pm$ 0.0108            &\multicolumn{1}{c}{...}         &  1.1277 $\pm$ 0.0088          & 1.2928 $\pm$ 0.0037   & 0.9408 $\pm$ 0.0202   \\
NGC 456 A No.3$^{\rm g}$       & 1.2134 $\pm$ 0.0117            &\multicolumn{1}{c}{...}         &\multicolumn{1}{c}{...}        &\multicolumn{1}{c}{...}& 0.9413 $\pm$ 0.0616   \\
\end{longtable}


\setcounter{figure}{0}


\begin{figure*}[t]
\vspace{-4.cm}
\vspace{-7.cm}
\caption{Optical images of the galaxies. The straight lines indicate the
location of the slit during observations (see Table~\ref{tab1}.)}
\label{fig1}
\end{figure*}


\begin{figure*}[t]
\caption{Flux-calibrated and redshift-corrected UVES 
spectra of galaxies.}
\label{fig2}
\end{figure*}


\begin{figure*}[t]
\caption{Flux-calibrated and redshift-corrected FORS low-resolution 
spectra of galaxies.}
\label{fig3}
\end{figure*}


\begin{figure*}[t]
\caption{Flux-calibrated and redshift-corrected FORS medium-resolution 
spectra of galaxies.}
\label{fig4}
\end{figure*}

\end{document}